\documentclass[useAMS,usenatbib]{mn2e} %
\pdfoutput=0

\usepackage{amsmath,amssymb,graphicx}
\usepackage{bm,color}
\usepackage{epstopdf}
\usepackage{times}

\graphicspath{{figs/}}

\bibpunct{(}{)}{;}{a}{}{,}

\newcommand{\dd}{{\mathrm d}}
\newcommand{\J}{{\mathrm J}}

\renewcommand{\vec}[1]{\boldsymbol{#1}}
\newcommand{\vt}{\vartheta}
\newcommand{\ve}{\varepsilon}

\newcommand{\ontop}[2]{
  \renewcommand{\arraystretch}{0.2}
  \begin{array}{c}
  #1 \\ #2
  \end{array}
  \renewcommand{\arraystretch}{1.0}
}
\newcommand{\lsim}{\ontop{<}{\sim}}

\newcommand{\Omegam}{\Omega_{\rm m}}

\newcommand{\Omegab}{\Omega_{\rm b}}

\newcommand{\mat}[1]{\textbfss{#1}}

\title
[CFHTLenS: cosmological model comparison using 2D weak lensing]
{CFHTLenS: Combined probe cosmological model comparison using 2D weak
  gravitational lensing}

\author[Kilbinger et al.]
{
  \parbox[h]{\textwidth}
  {
    Martin Kilbinger$^{1,2,3,4}$\thanks{E-mail:
      martin.kilbinger@cea.fr},
    Liping Fu$^5$,
    Catherine Heymans$^6$,
    Fergus Simpson$^6$,
    Jonathan Benjamin$^7$,
    Thomas Erben$^8$,
    Joachim Harnois-D\'eraps$^{9, 10}$,
    Henk Hoekstra$^{11, 12}$,
    Hendrik Hildebrandt$^{7, 8}$,
    Thomas D.~Kitching$^6$,
    Yannick Mellier$^{4, 1}$,
    Lance Miller$^{13}$,
    Ludovic Van Waerbeke$^7$,
    Karim Benabed$^4$,
    Christopher Bonnett$^{14}$,
    Jean Coupon$^{15}$,
    Michael J. Hudson$^{16, 17}$,
    Konrad Kuijken$^{11}$,
    Barnaby Rowe$^{18, 19}$,
    Tim Schrabback$^{8, 11, 20}$,
    Elisabetta Semboloni$^{11}$,
    Sanaz Vafaei$^{7}$,
    Malin Velander$^{13, 11}$
}
  \vspace*{10pt} \\
  \hspace{-.1cm}$^1$ CEA/Irfu/SAp Saclay, Laboratoire AIM, 91191 Gif-sur-Yvette, France\\
  \hspace{-.1cm}$^2$ Excellence Cluster Universe, Boltzmannstr. 2, 85748 Garching, Germany\\
  \hspace{-.1cm}$^3$ Universit\"ats-Sternwarte M\"unchen,
          Scheinerstr. 1, 81679 M\"unchen, Germany \\
  \hspace{-.1cm}$^4$ Institut d'Astrophysique de Paris, UMR7095 CNRS,
           Universit\'e Pierre \& Marie Curie, 98 bis boulevard Arago, 75014 Paris,
           France \\
   \hspace{-.1cm}$^5$ Key Lab for Astrophysics, Shanghai Normal
           University, 100 Guilin Road, 200234, Shanghai, China \\
   \hspace{-.1cm}$^6$ Scottish Universities Physics Alliance,
           Institute for Astronomy, University of Edinburgh, Royal
           Observatory, Blackford Hill, Edinburgh, EH9 3HJ, UK \\
   \hspace{-.1cm}$^7$ University of British Columbia, Department of
           Physics and Astronomy, 6224  Agricultural Road, Vancouver, B.C. V6T
           1Z1, Canada \\
   \hspace{-.1cm}$^{8}$ Argelander-Institut f\"ur Astronomie, Universit\"at Bonn,
        Auf dem H{\"u}gel 71, 53121 Bonn, Germany \\
   \hspace{-.1cm}$^9$ Canadian Institute for Theoretical Astrophysics, University of Toronto, M5S 3H8, Ontario, Canada \\
   \hspace{-.1cm}$^{10}$ Department of Physics, University of Toronto, M5S 1A7, Ontario, Canada \\
   \hspace{-.1cm}$^{11}$ Leiden Observatory, Leiden University, Niels
           Bohrweg 2, 2333 CA Leiden, The Netherlands \\
   \hspace{-.1cm}$^{12}$ Department of Physics and Astronomy, University of Victoria, Victoria, BC V8P 5C2, Canada \\
   \hspace{-.1cm}$^{13}$ Dept. of Physics, Oxford University, Keble Road, Oxford OX1 3RH, UK \\
   \hspace{-.1cm}$^{14}$ Institut de Ciencies de l’Espai, CSIC/IEEC, F. de Ciencies, Torre C5 par-2, Barcelona 08193, Spain \\
   \hspace{-.1cm}$^{15}$ Institute of Astronomy and Astrophysics, Academia Sinica, P.O. Box 23-141, Taipei 10617, Taiwan \\
   \hspace{-.1cm}$^{16}$ Department of Physics and Astronomy, University of Waterloo, Waterloo, ON, N2L 3G1, Canada \\
   \hspace{-.1cm}$^{17}$ Perimeter Institute for Theoretical Physics, 31 Caroline Street N, Waterloo, ON, N2L 1Y5, Canada \\
   \hspace{-.1cm}$^{18}$ Department of Physics and Astronomy, University College London, Gower Street, London WC1E 6BT, UK \\
   \hspace{-.1cm}$^{19}$ California Institute of Technology, 1200 E California Boulevard, Pasadena CA 91125, USA \\
   \hspace{-.1cm}$^{20}$ Kavli Institute for Particle Astrophysics and Cosmology, Stanford University, 382 Via Pueblo Mall,
                       Stanford, CA 94305-4060, USA \\
}

\begin{document}

\date{\today}

\pagerange{\pageref{firstpage}--\pageref{lastpage}} \pubyear{2009}

\maketitle

\label{firstpage}

\begin{abstract}

  We present cosmological constraints from 2D weak gravitational
  lensing by the large-scale structure in the Canada-France Hawaii
  Telescope Lensing Survey (CFHTLenS) which spans 154 square degrees
  in five optical bands. Using accurate photometric redshifts and
  measured shapes for 4.2 million galaxies between redshifts of 0.2
  and 1.3, we compute the 2D cosmic shear correlation function over
  angular scales ranging between
  0.8 and 350 arcmin. Using non-linear models of the dark-matter power
  spectrum, we constrain cosmological parameters by exploring the
  parameter space with Population Monte Carlo sampling. The best
  constraints from lensing alone are obtained for
  the small-scale density-fluctuations amplitude $\sigma_8$ scaled
  with the total matter density $\Omegam$.  For a flat $\Lambda$CDM
  model we obtain $\sigma_8 ( \Omegam / 0.27 )^{0.6} = 0.79 \pm
  0.03$.

  We combine the CFHTLenS data with WMAP7, BOSS and an HST
  distance-ladder prior on the Hubble constant to get joint
  constraints. For a flat $\Lambda$CDM model, we find $\Omegam = 0.283
  \pm 0.010$ and $\sigma_8 = 0.813 \pm 0.014$.
  In the case of a curved $w$CDM
  universe, we obtain $\Omegam = 0.27 \pm 0.03$, $\sigma_8 =
  0.83 \pm 0.04$, $w_0 = -1.10 \pm 0.15$ and $\Omega_K =
  0.006^{+0.006}_{-0.004}$.

  We calculate the Bayesian evidence to compare flat and curved
  $\Lambda$CDM and dark-energy CDM models. From the combination of all
  four probes, we find models with curvature to be at moderately
  disfavoured with respect to the flat case. A simple dark-energy
  model is indistinguishable from $\Lambda$CDM. Our
  results therefore do not necessitate any deviations from the
  standard cosmological model.

\end{abstract}


\section{Introduction}
\label{sec:intro}

Weak gravitational lensing is considered to be one of the most powerful tools of
cosmology. Its ability to measure both the geometry of the Universe and the
growth of structure offers great potential to obtain constraints on dark
energy and modified gravity. Moreover, to first order, weak lensing does
not rely on the relation between galaxies and dark matter (bias), and is
therefore a key probe of the dark Universe.

Cosmic shear, denotes the distortion of
images of distant galaxies due to the continuous deflection of light
bundles propagating through the inhomogeneous Universe. The induced
correlations between shapes of galaxies are directly related to the
statistical properties of the total (dark + luminous) large-scale
matter distribution. With an estimate of the redshift distribution of
the lensed galaxies, theoretical predictions of weak-lensing
observables can be tested to obtain constraints on cosmological
parameters and models. Recent reviews which also summarize past
observational results are \citet{BS01, vWM03,
  2008PhR...462...67M, 2008ARNPS..58...99H, 2010CQGra..27w3001B}.


The Canada-France Hawaii Telescope Legacy
Survey\footnote{\texttt{http://www.cfht.hawaii.edu/Science/CFHTLS}}
(CFHTLS) is a large imaging survey, offering a unique combination of
depth ($i_{AB} \lsim 24.5 $ at $5\sigma$ point source limiting
magnitude) and area (154 square degrees). It is the largest survey
volume over which cosmic shear has ever been measured. This paper
presents the first cosmological analysis of the complete CFHT Legacy
Survey with weak gravitational lensing. We measure the seond-order
cosmic-shear functions from the CFHTLenS (CFHT Lensing
Survey\footnote{\texttt{http://www.cfhtlens.org}}), which comprises
the final CFHTLS data. Earlier analyses of the CFHTLS used the first
data release (T0001) with 4 square degrees of the Deep survey
\citep{CFHTLSdeep} and 22 square degrees of the Wide part of the
survey \citep{CFHTLSwide}, followed by the third data release (T0003)
comprising 55 square degrees (\citeauthor{JonBen07}
\citeyear{JonBen07}; \citeauthor{FSHK08} \citeyear[hereafter
F08]{FSHK08}). The T0003 lensing data was subsequently employed in further
studies \citep{DMMK08, TSUK09, KB09}. Only CFHTLS-Wide $i^\prime$-band
data were used for those lensing analyses, and the redshift
distribution was inferred from the photometric redshifts from the Deep
survey \citep{2006A&A...457..841I}. Photometric redshifts in the Wide were
obtained subsequently with the T0004 release \citep{CIK09}. The
current series of papers uses the final CFHTLenS data release of 154
square degrees in the five optical bands $u^\ast, g^\prime, r^\prime,
i^\prime, z^\prime$. The analysis improved significantly in several
ways:

\begin{itemize}

\item The CFHTLS data have been reanalysed with a new 
  pipeline \citep{2009A&A...493.1197E, CFHTLenS-data}.

\item Photometric redshifts have been obtained for each individual
  galaxy in the lensing catalogue from PSF-homogenized images
  \citep{CFHTLenS-photoz}. The accuracy has been verified in detail by
  an angular cross-correlation technique \citep{CFHTLenS-2pt-tomo}.

\item PSF modeling and galaxy shape measurement have been performed with the
  forward model-fitting method \emph{lens}fit, which has
  been thoroughly tested on simulations and improved for CFHTLenS \citep{CFHTLenS-shapes}.

\item Systematics tests have been performed in a blind way, to yield
  unbiased cosmological results \citep{CFHTLenS-sys}.

\item The cosmology-dependent covariance matrix is obtained by a
  mixture of numerical simulations on small scales and analytical
  predictions on large scales.

\end{itemize}

The reliability and accuracy of our photometric redshifts allows for
three-dimensional weak lensing analyses. The measurement of the
lensing correlations for different redshift combinations allows us to
obtain information on the growth of structure
\cite[e.g.~][]{1999ApJ...522L..21H}, and has a great potential to
constrain dark energy and modified gravity models \citep[e.g.][]{DETF,
  2010GReGr..42.2219U}. {In several companion papers, we perform
  3D cosmic shear analyses by splitting up galaxies into redshift bins
  using correlation function methods presented here
  \citep[lensing tomography:][]{CFHTLenS-mod-grav, CFHTLenS-IA,
    CFHTLenS-2pt-tomo}}.

In this paper, we perform a 2D lensing analysis using a single
redshift distribution. Despite the fact that the redshift information
is not used in an optimal way, our analysis has several
advantages. First, it yields the highest signal-to-noise ratio ($S/N$)
for a single measurement. This is particularly important on large
angular scales, where the $S/N$ is too low to be used for
tomography. These large scales probe the linear regime, where
non-linear and baryonic effects do not play a role, and one can
therefore obtain very robust constraints on cosmology
\citep{2011MNRAS.417.2020S}. 
Second, we can include low-redshift
galaxies without having to consider intrinsic alignments
\citep[IA;][]{2004PhRvD..70f3526H}. For a broad redshift distribution,
IA is expected to be a sub-dominant contribution to the cosmological
shear-shear correlation with an expected bias for $\sigma_8$ which is
well within the statistical uncertainty \citep{2010MNRAS.408.1502K,
  2011MNRAS.410..844M, 2011A&A...527A..26J}, see also a joint lensing
and IA tomography analysis over the full available redshift range
\citep{CFHTLenS-IA}. Therefore, despite the fact that a 2D lensing
is more limited than tomography, it is less noisy and more
immune to primary astrophysical systematics. {It is therefore a
  necessary basic step} and puts any further
cosmological exploitation of CFHTLenS {using more advanced tomographic or
  full 3D lensing techniques} on solid grounds. {Such analyses
  are presented in the CFHTLenS companion papers.}

This paper is organized as follows. Section \ref{sec:cs} provides the
expressions for the second-order lensing observables used in this
analysis, both obtained from theoretical predictions and estimated
from data. The measured shear functions and covariances are presented
in Sect.~\ref{sec:cfhtlens_data}. Cosmological models and sampling
methods are introduced in Sect.~\ref{sec:setup}. The results on
cosmological parameters and models are presented in
Sect.~\ref{sec:results}, followed by consistency tests in
Sect.~\ref{sec:tests}.  The paper is concluded with a discussion in
Sect.~\ref{sec:discussion}.

\section{Weak cosmological lensing}
\label{sec:cs}

In this section the main relations between second-order weak lensing
observables and cosmological quantities are given. See
\cite{2010CQGra..27w3001B, 2008ARNPS..58...99H} for recent reviews.

\subsection{Theoretical background}
\label{sec:cs_theory}

Weak lensing by the large-scale structure measures
the convergence power spectrum $P_\kappa$, which can be
related to the total matter power spectrum $P_\delta$ via a projection
using Limber's equation \citep{1992ApJ...388..272K}:
\begin{equation}
  P_\kappa(\ell) = \int_0^{\chi_{\rm lim}} \dd \chi \,  G^2(\chi) \,
    P_\delta\left(k = \frac{\ell}{f_K(\chi)}; \chi\right).
\end{equation}
The projection integral is carried out over comoving distances $\chi$, from the
observer out to the limiting distance $\chi_{\rm lim}$ of the
survey. The lens efficiency $G$ is given by
\begin{equation}
  G(\chi) = \frac 3 2 \left(\frac{H_0} c \right)^2
  \frac{\Omegam}{a(\chi)} \int_\chi^{\chi_{\rm lim}} \dd \chi^\prime p(\chi^\prime)
  \frac{f_K(\chi^\prime - \chi)}{f_K(\chi^\prime)},
\end{equation}
where $H_0$ is the Hubble constant, $c$ the speed of light, $\Omegam$
the total matter density, and $a(\chi)$ the scale factor at comoving
distance $\chi$. The comoving angular distance is denoted with $f_K$
and depends on the curvature $K$ of the Universe;
{
\begin{equation}
f_K(w) = \left\{ \begin{array}{lll} 
      K^{-1/2} \sin{\left( K^{1/2} w \right) } \;\;\;\;\;\;
      & \mbox{for} \;\; & K>0 \\
      w & \mbox{for} & K=0 \\
      (-K)^{-1/2} \sinh{\left( (-K)^{1/2} w \right)} & \mbox{for} &
      K<0 \; .
    \end{array}
  \right.
\end{equation}
}
The 3D power
spectrum is evaluated at the wave number $k = \ell / f_K(\chi)$, where
$\ell$ denotes the projected 2D wave mode.  The function $p$
represents the weighted distribution of source galaxies.

\subsection{Flavours of real-space second-order functions}
\label{sec:2pt}

From an observational point of view, the most direct measurement of
weak cosmological lensing is in real space, by using the weak
gravitational shear signal as derived from galaxy ellipticity
measurements. The two-point shear correlation functions (2PCFs)
$\xi_+$ and $\xi_-$ are estimated in an unbiased way by averaging over
pairs of galaxies \citep{SvWKM02},
\begin{equation}
  \hat \xi_\pm(\vartheta) = \frac{\sum_{ij} w_i w_j [\varepsilon_{\rm
      t}(\vec \vartheta_i) \, \varepsilon_{\rm t}(\vec \vartheta_j) \pm
    \varepsilon_{\times}(\vec \vartheta_i) \, \varepsilon_{\times}(\vec \vartheta_j)]}
  {\sum_{ij} w_i w_j}.
\label{xipmestim}
\end{equation}
The sum is performed over all galaxy pairs $(ij)$ with angular
distance $|\vec \vt_i - \vec \vt_j|$ within some bin around
$\vartheta$.  With $\varepsilon_{\rm t}$ and $\varepsilon_\times$ we
denote the tangential and cross-component of the galaxy ellipticity,
respectively.  The weights $w_i$ are obtained from the
\emph{lens}fit shape measurement pipeline \citep{CFHTLenS-shapes}.
The 2PCFs are the Hankel transforms of the convergence power spectrum
$P_\kappa$ or, more precisely, of linear combinations of the E- and
B-mode spectra, $P_{\rm E}$ and $P_{\rm B}$, respectively. Namely,
\begin{align}
  \xi_\pm(\vartheta)
  & =  {1\over 2\pi} \int_0^\infty {\rm d}\ell\,  \ell \, \left[ P_{\rm
      E}(\ell) \pm P_{\rm B}(\ell) \right] {\rm J}_{0,4}(\ell\vartheta),
  \label{xipmpkappa}
\end{align}
where ${\rm J}_0$ and ${\rm J}_4$ are the first-kind Bessel functions of
order 0 and 4, and correspond to the components $\xi_+$ and $\xi_-$,
respectively.

It is desirable to obtain observables which only depend on the E-mode
and B-mode, respectively. Weak gravitational lensing, to first order,
only gives rise to an E-mode power spectrum and therefore, a
non-detection of the B-mode is an important sanity check of the
data. To this end, we calculate the following second-order shear
quantities which can be derived from the correlation functions: The
aperture-mass dispersion $\langle M_{\rm ap}^2 \rangle$
\citep{1998MNRAS.296..873S}, the shear top-hat rms $\langle |\gamma|^2
\rangle$ \citep{1992ApJ...388..272K}, the optimized ring statistic
${\cal R}_E$ \citep{FK10}, and COSEBIs \citep[Complete Orthogonal Sets
of E-/B-mode Integrals;][]{COSEBIs}. The optimized
ring statistic was introduced as a generalisation of the so-called
ring statistic \citep{SK07}. The corresponding filter functions have
been obtained to maximise the figure-of-merit of $\Omega_{\rm m}$ and
$\sigma_8$ for a CFHTLS-T0003-like survey. We use these functions for
CFHTLenS which, despite the larger area, has similar survey
characteristics. COSEBIs represent yet another generalisation and
contain all information about the E- and B-mode weak-lensing field
from the shear correlation function on a finite angular range.

Being quantities obtained from the 2PCFs by non-invertible relations,
these derived functions do not contain the full information about the
convergence power spectrum \citep{EKS08}, but separate the E- and the
B-mode in a more or less pure way, as will now be described.
The derived second-order functions can be written as integrals
over the filtered correlation functions. They can be estimated as follows:
\begin{equation}
X_{\rm E, B} = \frac 1 2 \sum_i \vt_i \, \Delta \vt_i
\left[ F_+\left( {\vt_i} \right) \xi_+(\vt_i) \pm
       F_-\left( {\vt_i} \right) \xi_-(\vt_i) \right] .
     \label{X_EB}
\end{equation}
Here, $\Delta \vartheta_i$ is the bin width, which can vary with $i$, for
example in the case of logarithmic bins. With suitable filter functions
$F_+$ and $F_-$ {(App.~\ref{sec:filter_functions})}, the estimator $X_{
\rm E}$ ($X_{\rm B}$) is sensitive to the
E-mode (B-mode) only. The filter functions are defined for the various
second-order observables in Table \ref{tab:F+-} and Appendix
\ref{sec:filter_functions}.

All derived second-order functions are calculated for a family of
filter functions. For the aperture-mass dispersion, the optimized
ring statistic, and the top-hat shear root mean square (rms),
these are given for a continuous parameter $\theta$ which can be
interpreted as the smoothing scale. Here and in the following we will
use the notation '$\vartheta$' as the scale for the 2PCFs, and
'$\theta$' as the smoothing scale for derived functions.

For COSEBIs, the filter functions
are a discrete set of functions. The latter exist in two flavours,
Lin-COSEBIs and Log-COSEBIs, defined through filter functions $F_\pm$
which are polynomials on linear and logarithmic angular scales,
respectively. Here we use Log-COSEBIs, for which many fewer
modes are required to capture the same information as Lin-COSEBIs
\citep{COSEBIs, 2012A&A...542A.122A}. {See App.~\ref{sec:filter_functions}
for more details}.

All of the above functions can be expressed in terms of the
convergence power spectrum. The general relation is
\begin{align}
  X_{\rm E, B} = \frac 1 {2\pi} \int_0^\infty \dd \ell \, \ell \,
  P_{\rm E, B}(\ell) \hat U^2(\ell) .
\label{X_EB_Fourier}
\end{align}

The functions $F_{\pm}$ and $\hat U^2$ are Hankel-transform pairs,
their relation is given by \citet{2002ApJ...568...20C} and 
\citet{2002A&A...389..729S} as
\begin{equation}
  F_\pm(x) = \int_0^\infty \dd t \, t \, \J_{0,4}(x t) \hat
  U^2(t).
  \label{T_pm}
\end{equation}

\begin{table*}

  \caption{E- and B-mode separating second-order functions. They are
    estimated as integrals, or sums in the discrete case, over the
    2PCFs $\xi_\pm$ multiplied with the filter
    functions $F_{\pm}$ with the formal integration range
    $[\vt_{\rm min}; \vt_{\rm max}]$, see eq.~(\ref{X_EB}). The argument $\vt$
    denotes the integration variable.
  }
  \label{tab:F+-}

  \begin{tabular}{lcccccl}
    Name & $X_{\rm E, B}$ & $F_{\pm}$ (eq.~\ref{X_EB}) & $\hat U$ (eq.~\ref{X_EB_Fourier}) & $\vt_{\rm
      min}$ & $\vt_{\rm max}$ & Reference \\ \hline
    Aperture-mass dispersion & $\langle M_{\rm ap,
      \times}^2\rangle(\theta = \vt_{\rm max} / 2)$ &
    $T_\pm(\vt / \theta) / \theta^2$ (eq.~\ref{T}) &
    eq.~(\ref{Uhat-map}) & 0 & $2 \theta$ &
    \cite{1998MNRAS.296..873S} \\
    Top-hat shear rms & $\langle | \gamma |^2 \rangle(\theta)$ &
    $S_\pm(\vt / \theta) / \theta^2$ (eq.~\ref{S}) &
    eq.~(\ref{Uhat-gamma}) & 0 &
    $\infty$ & \cite{1992ApJ...388..272K} \\
    Optimized ring statistic & ${\cal R}_{\rm E, B}(\theta = \vt_{\rm max})$ &
    $T_{\pm}(
    \vt)$ (eq.~\ref{TEB}) &
    n/a & $\vt_{\rm min}$ & $\vt_{\rm max}$ & \cite{FK10} \\
    COSEBIs & $E_n, B_n$ & $T^{\rm log}_{\pm, n}(\vt)$ (eq.~\ref{t}) & n/a & $\vt_{\rm min}$
    & $\vt_{\rm max}$ & \cite{COSEBIs} \\ \hline\hline
    \end{tabular}

\end{table*}

\subsubsection{Finite support}

The measured shear correlation function is available only on a finite
interval $[\vt_{\rm min}; \vt_{\rm max}]$.  The upper limit is given
by the finite survey size. We choose $460$ arc minutes, which is
roughly the largest scale for which a sufficient number of galaxy
pairs are available on at least three of the four Wide
patches\footnote{W2 as the smallest patch extends to 400 arc minutes,
  whereas W1 probes scales as large as 685 arc minutes.}. The lower
limit comes from the fact that for close galaxy pairs, shapes cannot
be measured reliably. For \emph{lens}fit, galaxies with separations
smaller than the postage stamp size of $48$ pixels $\simeq 9$ arcsec
tend to have a shape bias along their connecting direction
\citep{CFHTLenS-shapes}. Since this direction is randomly orientated
to a very good approximation, we do not have to remove those galaxies
altogether; setting a minimum angular separation of $\vt_{\rm min} =
9$ arcsec does avoid the correlation of those pairs. Both galaxies
from a close pair can be independently correlated with other, more
distant galaxies, for which the close-pair shape bias acts as a
second-order effect and can safely be neglected.

If the support of the filter functions $F_+$ and $F_-$ exceeds the
observable range, eq.~(\ref{X_EB}) leads to biased results, and a pure
E- and B-mode separation is no longer guaranteed.

This is the case for the aperture-mass dispersion and the top-hat
shear rms. For the former, only the lower angular limit is problematic
and causes leakage of the B-mode into the E-mode signal on small
smoothing scales. On scales larger than $\theta_{\rm min} = 5.5$ arc
minutes however, this leakage is below 1.5 cent for $\vt_{\rm
  min} = 9$ arcsec \citep{KSE06}. We therefore choose $\theta_{\rm
  min} = 5.5$ arcmin to be the smallest smoothing scale for $\langle
M_{\rm ap}^2\rangle(\theta)$. Note that on scales smaller than
$\vt_{\rm min}$ we set the correlation function to zero and do not use
a theoretical model to extrapolate the data on to this range, to avoid
a cosmology-dependent bias.

The B-mode leakage for the top-hat shear rms $\left\langle | \gamma^2 |
\right\rangle$ is a function of both the lower and upper available
angular scales. Over our range of scales, the predicted leaked B-mode
for the WMAP7 cosmology is nearly constant with a value of $5.3 \times
10^{-7}$.

The first pure E-/B-mode separating function for which the
corresponding filter functions have finite support was introduced in
\cite{SK07}.  Following that approach, the optimized ring statistic
and COSEBIs were constructed in a similar way to not suffer
from an E-/B-mode leakage.

An additional bias arises from the removal of close galaxy pairs in
the lensing analysis, as was first reported by
\cite{2011A&A...528A..51H}. As previously discussed, \emph{lens}fit
produces a shape bias for galaxies separated by less than $9$ arcsec,
but this bias is random and the close pairs are therefore used in the
analysis, to be correlated with other galaxies at larger
distances. That said, for very close blended galaxies, where it is
non-trivial to determine if there is one or two galaxies observed,
galaxy shapes cannot even be attempted. These blended pairs
are therefore not reliably detected and lost from our analysis.  This
causes a potential bias, since these galaxies are removed
preferentially at low redshift, where galaxy sizes are larger,
and from high-density regions compared to voids, because galaxies
trace the large-scale structure.  From Fig.~3 of
\citet{2011A&A...528A..51H} we infer that the magnitude of this
effect is at the per cent level on scales larger than $0.8$ arcmin
for the 2PCFs.

\section{CFHTLenS shear correlation data and covariance}
\label{sec:cfhtlens_data}

The CFHTLenS data are described in several companion papers; for a
full summary see \cite{CFHTLenS-sys}. Stringent systematics tests have
been performed in \cite{CFHTLenS-sys} which flag and remove any data
in which significant residual systematics are detected. It is this
cleaned sample, spanning $\sim 75$ per cent of the total CFHTLenS
survey area, that we use in this
analysis.
This corresponds to 129 out of 171 MegaCam pointings.
In this paper, we complement those tests by measuring
the B-mode up to large scales (Sect.~\ref{sec:results-EB}). Comparing
this to the previous analysis of F08 shows the improved
quality of the lensing analysis by CFHTLenS.

\subsection{Redshift distribution}
\label{sec:nofz}

\begin{figure}  
  \resizebox{\hsize}{!}{
    \includegraphics[bb=50 50 395 302]{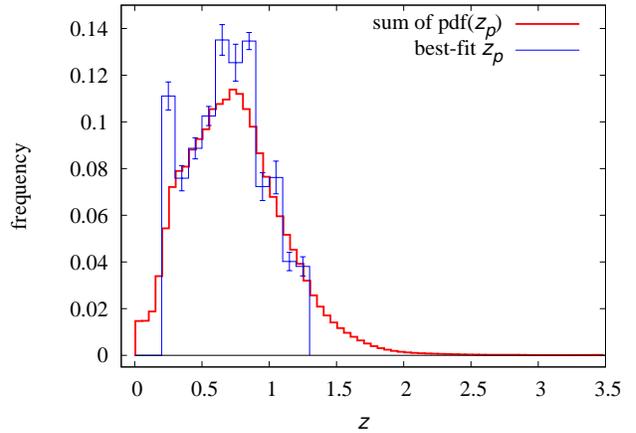}
  }
  
  \caption{The redshift distribution $p(z_{\rm p})$ histogram, estimated as
    the sum of pdfs for $0.2 < z_{\rm p} < 1.3$ (thick red curve). For
    comparison, the histogram of the best-fitting
    photo-$z$s is shown as thin blue curve. Error bars correspond to
    the variance between the four Wide patches.}
  \label{fig:nofz}
\end{figure}

A detailed study of the reliability of our photometric redshifts, 
the contaminations between redshift bins, and the cosmological
implications is performed in \cite{CFHTLenS-2pt-tomo}. This work shows
that the true redshift distribution $p(z)$ is well approximated by the
sum of the probability distribution functions (pdfs) for all
galaxies. The pdfs are output by BPZ \citep[Bayesian Photometric
Redshift Estimation;][]{2000ApJ...536..571B} as a function of
photometric redshift $z_{\rm p}$, and have been obtained by
\citet{CFHTLenS-photoz}. The resulting $p(z)$ is consistent with the
contamination between redshift bins as estimated by an angular
cross-correlation analysis \citep{BvWMK10}. The contamination is
relatively low for galaxies selected with $0.2 < z_{\rm p} < 1.3$,
which is confirmed by a galaxy-galaxy-lensing redshift scaling
analysis in \cite{CFHTLenS-sys}. The resulting $p(z)$ is shown in
Fig.~\ref{fig:nofz}. The mean redshift is $\bar z = 0.748$. In
contrast, the mean redshift of the best-fitting $z_{\rm p}$ histogram is
biased low with $\bar z = 0.69$.

\subsection{Angular correlation functions}
\label{sec:xi}

We calculate the two-point shear correlation functions by averaging
over pairs of galaxies, using the tree code
\textsc{athena}\footnote{\texttt{http://www2.iap.fr/users/kilbinge/athena}}. Galaxies
are partitioned into nested branches of a tree, forming rectangular
boxes in right ascension $\alpha$ and declination $\delta$. For two
branches at angular distance $\vartheta$ and box sizes $d_i$ ($i=1, 2$
along the $\alpha$ and $\delta$ direction, respectively) if both
opening angles $\omega_i \equiv d_i / \vartheta$ are smaller than a
threshold angle $\omega_{\rm th}$, the tree is not followed further
down by descending into sub-branches. Instead, the weighted average
shear in each branch is used for the 2PCF estimator
(eq.~\ref{xipmestim}). We found that a value of $\omega_{\rm th} =
0.03$ gives sufficient accuracy compared to the brute-force approach.

We use distances and angles on the sphere to calculate the shear correlation
functions. For two galaxies  $i=1, 2$ at right ascension and declination
$(\alpha_i, \delta_i)$, we calculate the great-circle distance $\vartheta$ with
\begin{equation}
  \cos \vartheta = \cos(\alpha_2 - \alpha_1) \cos \delta_1 \cos \delta_2 + \sin \delta_1
  \sin \delta_2 .
  \label{great_circle_distance}
\end{equation}
Each galaxy's ellipticity
is measured in a
local Cartesian coordinate system with the $x$-axis going along the line of
constant declination and the $y$-axis pointing to the North pole. We
project this ellipticity to the tangential and radial component with
respect to the connecting great circle. For that, we calculate the
angle $\beta_i$ between the great circle segments $\vartheta$ and
$\alpha_i$. Then, the projection or so-called course angles are
$\phi_i = \pi/2 - \beta_i$. With the sine and cosine rules on the
sphere, we get
\begin{align}
  \cos \phi_1 = & \frac{\sin (\alpha_2 - \alpha_1) \cos \delta_2}{\sin \vartheta};
  \nonumber \\
  \sin \phi_1 = & \frac{\cos \delta_2 \sin \delta_1 - \sin \delta_2 \cos
    \delta_1 \cos (\alpha_2 - \alpha_1)}{\sin \vartheta},
  \label{geodesic_correction}
\end{align}
and corresponding expressions for $\phi_2$ by exchanging indices.

\begin{figure}

  \begin{center}
    \input{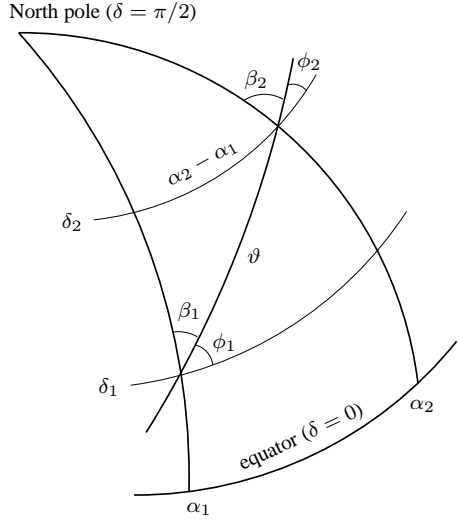}
    \end{center}

  \label{fig:angles}

  \caption{Angles and coordinates on a sphere for two galaxies $i=1,2$ located at $(\alpha_i, \delta_i)$. Great circle segments
    are drawn as bold lines.}

\end{figure}


To estimate the smoothed second-order quantities, we compute the 2PCFs
on 10,000 linear angular bins. This is large enough not to cause a
significant E-/B-mode leakage due to the approximation of the
integrals over the correlation functions by the direct sum
(eq.~\ref{X_EB}). We verified this using CFHTLenS numerical
simulations with no B-mode (see next Section) ,
see also \cite{2012arXiv1208.0068B}.  We choose the smallest angular
distance between two galaxies to be 9 arcsec, corresponding to
the first bin centre to be $10.4$ arcsec. We calculate the 2PCFs
(eq.~\ref{xipmestim}) as the weighted mean over the four Wide patches,
using the number of galaxy pairs as weight for each bin.

\subsection{Data covariance}
\label{sec:cov}

\begin{figure}
  
  \begin{center}
    \resizebox{1.1\hsize}{!}{
      \includegraphics[bb=155 155 445 640]{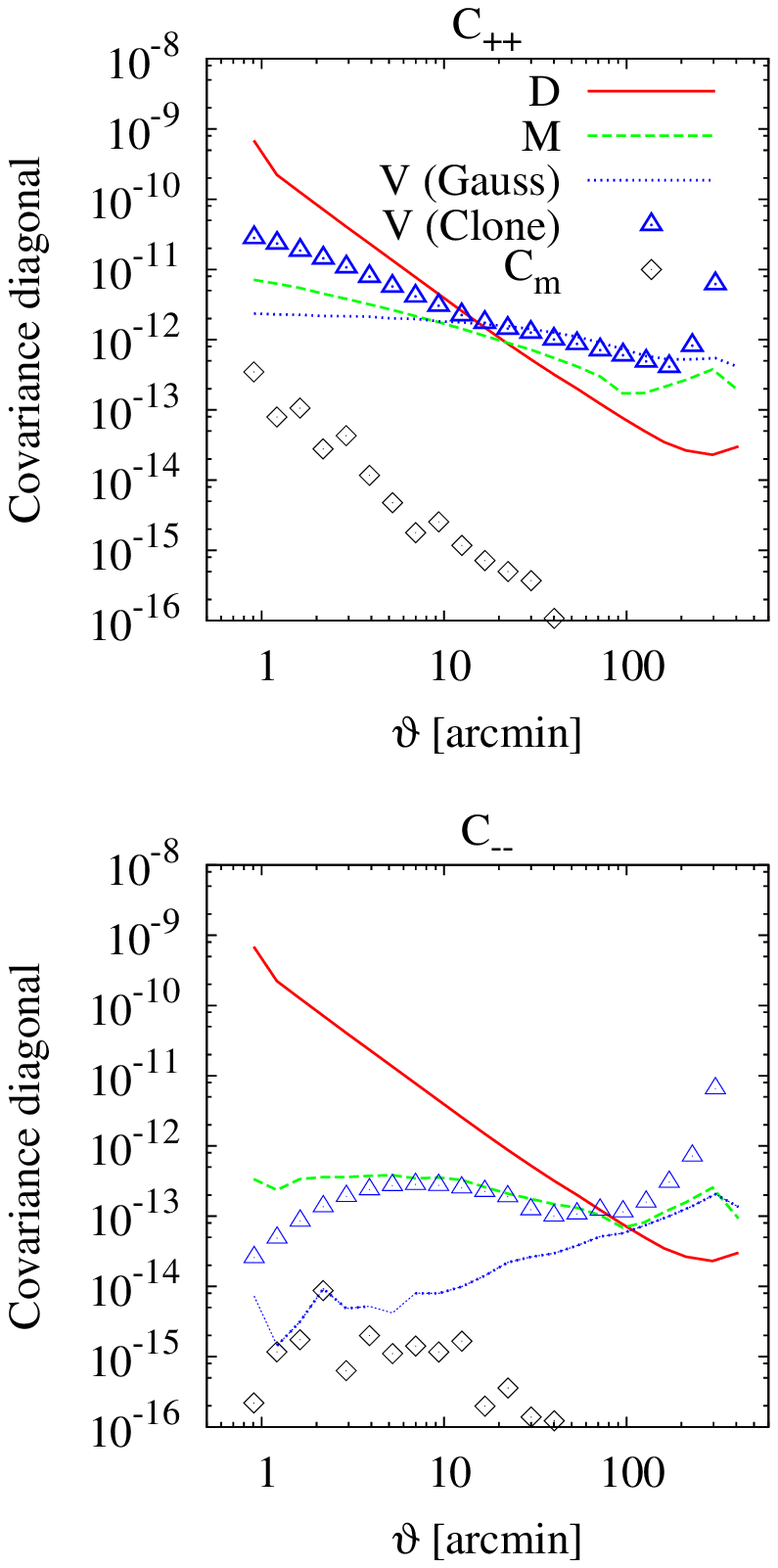}
    }
  \end{center}

  \caption{Diagonal of the covariance $\mat C_{++}$ (\emph{top panel}) and
    $\mat C_{--}$ (\emph{bottom}), split up into various terms: Shot-noise
    $\mat D$ (solid red line), mixed term $\mat M$ (dashed green), cosmic
    variance $\mat V$ (dotted blue line and crosses) and shear calibration
    covariance $\mat C_m$ (see Sect.~\ref{sec:calib}).}
  \label{fig:covdiag}
\end{figure}

\begin{figure}
  
  \begin{center}
    \resizebox{\hsize}{!}{
      \includegraphics{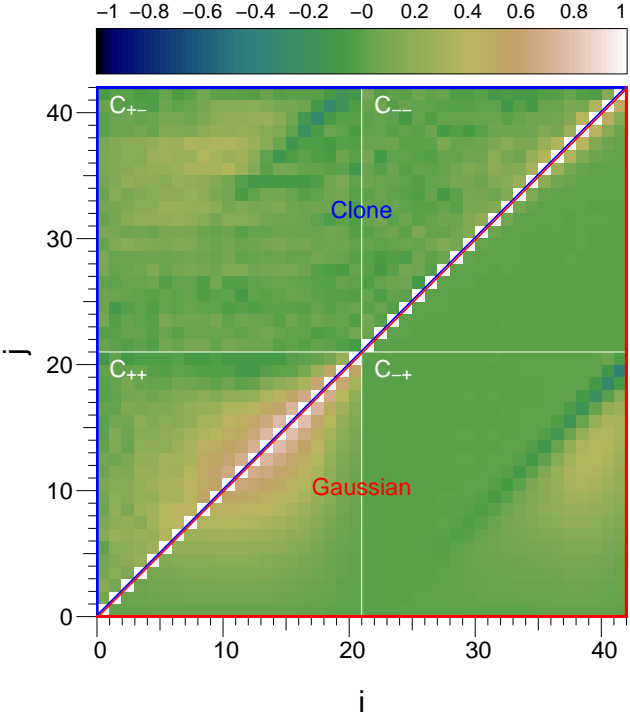}
    }
  \end{center}

  \caption{Correlation coefficient of the total covariance, shot-noise
    plus mixed plus cosmic variance. The lower right triangle (within
    the red boundary) is the Gaussian prediction, the upper left
    (blue) triangle shows the total covariance with non-Gaussian parts
    from the Clone. The four blocks are $\mat C_{+-}, \mat C_{--},
    \mat C_{++}$ and $\mat C_{-+}$, respectively, as indicated in the plot.
    The axes labels are the matrix indices $i$ and $j$.}
  \label{fig:rcovtot}
\end{figure}

To model and interpret the observed second-order shear functions, we need to
estimate the data covariance and its inverse. The cosmic shear covariance $\mat C$ is
composed of the shot-noise $\mat D$, which only appears on the
diagonal, a cosmic-variance contribution $\mat V$, and a mixed term
$\mat M$ \citep{SvWKM02}. The covariance of the 2PCFs is comprised of
four block matrices. The diagonal consists of $\mat C_{++}$ and $\mat
C_{--}$ which are the auto-correlation covariance matrices of $\xi_+$
and $\xi_-$, respectively. The off-diagonal blocks are $\mat C_{+-}$
and $\mat C_{-+} = \mat C_{+-}^{\rm t}$ which denote the
cross-correlation covariance between $\xi_+$ and $\xi_-$.

Since the cosmic shear field is
non-Gaussian on small and medium angular scales, the cosmic variance
involves four-point functions. Neglecting those can yield
overly optimistic cosmological constraints \citep{2007MNRAS.375L...6S, 2009MNRAS.395.2065T,
  2011A&A...536A..85H}.

To account for non-Gaussianity, we use $N$-body simulations from
\cite{CFHTLenS-Clone}. From these simulations, a `Clone' of the
CFHTLenS data has been produced with the same galaxy redshift
distribution, galaxy clustering, masks, and noise properties.
{The cosmological lensing signal is added using ray-tracing
through the light cones.}  The
Clone cosmology is a flat $\Lambda$CDM model with $\Omegam = 0.279,
\Omegab = 0.046, n_{\rm s} = 0.96, \sigma_8 = 0.817$ and $h = 0.701$.
The lensing signal for each galaxy is constructed by ray-shooting
through the simulated dark-matter distribution. Each simulated line of
sight spans {a field of view of} $3.5 \times 3.5$ square degrees.
We fit close to $4 \times
4$ MegaCam pointings on each line of sight, which is possible because
of overlapping areas between pointings. A total of 184 independent
lines of sight
are used to calculate the field-to-field covariance matrix. The final
matrix is scaled with the ratio of the effective areas (including
masks) of 0.11 which corresponds to 90 per cent of the area of 16
MegaCam pointing that fit into each line of sight, divided by 129
MegaCam pointings used in this analysis. We average over three
different samples of the galaxy redshift probability distribution,
where galaxy redshifts were drawn from the corresponding pdf.

As shown in the upper panel of Fig.~\ref{fig:covdiag}, the Gaussian
prediction for the cosmic variance \citep{KS04} for $\xi_+$ provide a
good match to the Clone covariance on intermediate scales, $10$ arcmin
$\lsim \vartheta \lsim 30$ arcmin. On larger scales, up to $\vartheta
< 200$ arcmin, the numerical simulations under-predict the power due
to the finite box size \citep[e.g.][]{2006MNRAS.370..691P}. Only the
last two data points show an increased variance, which is due to the
finite Clone field geometry. When comparing the Clone mean correlation
function to a theoretical prediction with cut-off scale $k = (2
\pi/147) \; h^{-1}\!$ Mpc, we get a rough agreement between the two,
indicating that the lack of power is indeed caused by the finite box.
We draw similar conclusions for the cosmic variance of $\xi_-$, shown
in the lower panel of Fig.~\ref{fig:covdiag}. Further, we verified
that a Jackknife estimate of the variance by sub-dividing the CHFLTenS
data into 129 subfields gives consistent results.

\subsubsection{Grafting the covariance matrix}
\label{sec:grafting}

We construct the total covariance out to $\vartheta = 350$ arcmin by
grafting the Clone covariance $\mat V_{\rm cl}$ to the analytical
Gaussian prediction. For the latter, we use the method developed in
\citet{KS04}, which takes into account the discrete nature of the
galaxy distribution and the field geometry. First, we add the Clone
covariance $\mat V_{\rm cl}$ to the Gaussian cosmic covariance term
$\mat V_{\rm G}$. The combined cosmic covariance is
\begin{equation}
  V_{ij} = g_{ij} V_{{\rm cl}, ij} + (1 - g_{ij}) V_{{\rm G}, ij},
  \label{cv_modulate}
\end{equation}
where the modulation function $g_{ij}$ alleviates discontinuities in
the combined matrix.
We choose $g_{ij}$ to be a bi-level step function, with $g_{s, s} =
1/2$; $g_{i j} = 1$ if both indices $i, j$ are smaller than the step
index $s$; and $g_{i j} = 0$ if at least one of the indices $i$ or $j$
is larger than or equal to $s$.
The step index $s$ is chosen such that $\vartheta_s$ is the scale closest
to 30 arcmin. Equation (\ref{cv_modulate}) is applied to all covariances
between the two shear correlation functions, i.e.~$\mat V_{++}, \mat V_{+-}$ and
$\mat V_{--}$.

The Clone covariance also contains {an additional variance term,
  which was discovered recently \citep{2009ApJ...701..945S}. This
  so-called \emph{halo sample variance} (HSV) stems from density
  fluctuations on scales larger than the (finite) survey size that are
  correlated with fluctuations on smaller scales. For example, the
  number of halos in the survey depend on the large-scale modes
  outside the survey, since halos are clustered and do not just follow
  a Poisson distribution. This introduces an extra variance to the
  measured power spectrum.  The halo sample variance is proportional
  to the rms density fluctuations at the survey scale
  \citep{2009ApJ...701..945S}.  Since our simulated light-cones are
  cut-outs from larger boxes of size $L = 147 \, {\rm Mpc}/h$ ($L =
  231 \, {\rm Mpc}/h$) at redshift below (above) unity, they do
  contain Fourier scales outside the survey volume and their coupling
  to smaller scales.}  The halo sample variance is important on small
scales, where our cosmic variance is dominated by the Clone
covariance. Following \citet{2009ApJ...701..945S}, we estimate the
halo sample variance to dominate the CFHTLenS total covariance at
$\ell \approx 2 \times 10^3$, corresponding to 5 arc minutes which is
the Clone covariance regime.

The missing large-scale Fourier modes in the simulation box cause the
HSV to be underestimated. A further underestimation comes from the
rescaling of the Clone lines of sight to the CFHTLenS area since, in
contrast to the other covariance terms, the HSV term
decreases less strongly than the inverse survey area
\citep{2009ApJ...701..945S}. According to \citet{2012arXiv1207.6322K},
when naively rescaling from a 25 deg$^2$-survey to 1500 deg$^2$, the
signal-to-noise ratio is too optimistic by not more than 10 per cent. For a
re-scaling to the smaller CFHTLenS area, this bias is expected to
be much less.


\subsubsection{Cosmology-dependent covariance}
\label{sec:cosmo-dep-cov}

{Our grafted covariance of the two-point correlation function is estimated
for a fiducial cosmological model, which is given by the $N$-body simulations.
In order not to bias the likelihood function of the data (Sect.~\ref{sec:sampling})
at points other than that fiducial model, we need to account for the fact that
the covariance depends on cosmological parameters.
}
We model the cosmology-dependence of the covariance matrix following
\cite{2009A&A...502..721E}{, who suggested approximative schemes for
the mixed term $\mat M$ and the cosmic variance term $\mat V$. Accordingly,}
for the cosmic-variance term, we assume a
quadratic scaling with the shear correlation function. This is true on
large scales, where {the shear field is close to Gaussian and}
the covariance is indeed
proportional to the square of the correlation function. We calibrate the
small-scale Clone covariance in the same way, as any differences in
the way the non-Gaussian part might scale are likely to be small.

For the mixed term $\mat M$, we use the fitting formula provided by
\citet{2009A&A...502..721E}. They approximate the variation with
$\Omegam$ and $\sigma_8$, leaving the matrix fixed for other
parameters. The shot-noise term $\mat D$ does not depend on cosmology. The
final expression for the covariance matrix is
\begin{align}
  C_{\mu\nu, ij}(\vec p) = 
  & \; D_i \delta_{ij}\delta_{\mu\nu} \nonumber \\
  & + M_{\mu\nu, ij}(\vec p_0) \left(
    \frac{\Omegam}{0.25} \right)^{\alpha(\mu, \nu, i, j)} \left(
    \frac{\sigma_8}{0.9} \right)^{\beta(\mu, \nu, i, j)} \nonumber \\
  & + V_{\mu\nu, ij}(\vec p_0)
  \frac{\xi_\mu(\vartheta_i, \vec p)}{\xi_\mu(\vartheta_i, \vec p_0)}
  \frac{\xi_\nu(\vartheta_j, \vec p)}{\xi_\nu(\vartheta_j, \vec p_0)},
  \label{cov_ESH}
\end{align}
where the indices $\mu, \nu$ stand for the components `$+$' and `$-$',
and $i, j$ are the angular-scale indices.  Here, $\vec p = (\Omegam,
\sigma_8, \ldots)$ denotes the cosmological parameter for which the
covariance is evaluated, with $\vec p_0$ being the fiducial model of
the Clone simulation (Sect. \ref{sec:cov}).  The shot-noise term $D_i$
and the mixed term $M_{\mu\nu, i,j}$ are estimated using the method of
\cite{KS04}. The cosmic-variance term $V_{\mu\nu, ij}$ is given in
eq.~(\ref{cv_modulate}).  The power-law indices $\alpha$ and $\beta$
depend on the angular scales and the covariance component $(\mu, \nu)
\in \{$`$+$'$, $`$-$'$\}$, and have been obtained in
\citet{2009A&A...502..721E}. We note that the inverse covariance is
very sensitive to two apparent outliers of $\alpha$ and $\beta$ for
the $\mat C_{+-}$-part. To avoid numerical issues, we replace these
two numbers by the mean of their neighbouring values.
{In our case the mixed term of the covariance is important and
cannot be neglected \cite[see also][]{KS04, 2010APh....32..340V},
in contrast to recent findings by \cite{2012arXiv1210.2732J}.}

During the Monte-Carlo sampling (Sect.~\ref{sec:sampling}), the covariance is
updated at each sample point $\vec p$ using eq.~(\ref{cov_ESH}). We
make sure that each calculation of the covariance resulted in a
numerically positive-definite matrix, and discard the (rare) sample
points for which this is not the case.

In Fig.~\ref{fig:rcovtot} we show the total covariance $\mat C = \mat
D + \mat M + \mat V$ and compare it to the Gaussian prediction $\mat
C_{\rm G} = \mat D + \mat M + \mat V_{\rm G}$.  Both cases are similar
on most scales. On small scales the grafted covariance shows stronger
cross-correlations between scales, indicating non-Gaussian effects.
We find that the additional covariance $\mat C_m$ due to the shear
calibration (see Sect.~\ref{sec:calib}) can be neglected, as can be
seen in Fig.~\ref{fig:covdiag} and Sect.~\ref{sec:sys}.

\subsubsection{Inverse covariance estimator}

It has been shown in \cite{andersen03} and \cite{HSS07} that the
maximum-likelihood (ML) estimator of the inverse covariance is biased
high. The field-to-field covariance from the Clone is such an ML
estimate. The bias depends on the number of realisations or fields
$n$, and the number of bins $p$. The ML estimator can be de-biased by
multiplication with the Anderson-Hartlap factor $\alpha = (n - p -
2)/(n - 1)$ \citep{HSS07}.

\begin{figure}
  
  \resizebox{1.0\hsize}{!}{
    \includegraphics[bb=65 49 370 303]{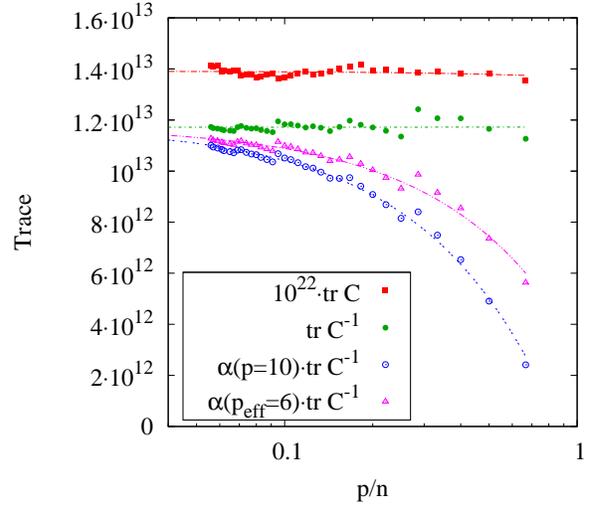}
  }
  
  \caption{Trace of the covariance and inverse grafted cosmic
    covariance, plotted against the ratio of 
    the number of bins $p$ to the number of realisations $n$.
    Shown are the cases for covariance (red squares), the inverse
    (green filled circles), and two cases where the inverse has been
    multiplied with the Anderson-Hartlap factor $\alpha$,
    corresponding to the number of bins $p=10$ (blue open circles),
    and $p=6$ (magenta triangles), causing an over-correction in both
    cases. This shows that we can use the inverse covariance estimator
    without correction (green curve).}
  \label{fig:hartlap}
\end{figure}

Our final 2PCF covariance, however, is the mixture of an ML estimate and
analytical expressions. The ML estimator is modulated with the
Gaussian cosmic variance via eq.~(\ref{cv_modulate}), to which we add
the shot-noise and mixed terms, eq.~(\ref{cov_ESH}).  We quantify a
possible bias of the inverse covariance $(\mat C_{++})^{-1}$ by
varying $n$ for a fixed $p = 10$. For a step index $s = 7$,
corresponding to $\vartheta_s = 37$ arcmin, Fig.~\ref{fig:hartlap}
shows that the trace of $(\mat C_{++})^{-1}$ does not depend on the
ratio $p/n$ for our grafted cosmic covariance matrix. Multiplication
with $\alpha$ for $p=10$ results in an overcorrection, causing a
strong decrease of ${\rm tr} (\mat C_{++})^{-1}$ with $p/n$. A similar
albeit less strong decrease is seen when naively taking into account
the fact that the Clone covariance only contributes to an effective
number of scales $p_{\rm eff} = 6$, according to
eq.~(\ref{cv_modulate}) with $s = 7$. The three curves for the inverse
seem to converge for $p/n \rightarrow 0$. Therefore, we have reason to
be confident that any bias of the unaltered inverse of eq.~(\ref{cov_ESH}) is
small, and hence we do not need to apply the scalar correction
factor $\alpha$. The addition of a deterministic component to the ML
covariance seems to be sufficient to render the estimate of the
inverse to be unbiased.

\subsubsection{Covariance of derived second-order functions}
\label{sec:cov-sm}

Expressions for the covariance of the derived second-order statistics
(eq.~\ref{X_EB}) are straight-forward to obtain, and can be calculated
by integrating the covariance of the 2PCFs \citep{SvWKM02}. However,
the necessary precision {for the numerical integration}
requires a large number of angular bins for
which the 2PCF covariance has to be calculated, which is very
time-consuming. Consequently, for all derived second-order functions
we choose not to graft the Clone covariance to the Gaussian
covariance, but instead only use the Clone to calculate the total
covariance of the derived functions. To include shot noise, we add to
each galaxys' shear an intrinsic ellipticity as a Gaussian random
variable with zero mean and dispersion $\sigma_{\varepsilon}$ = 0.38.
The latter is calculated as $\sigma^2_{\varepsilon} = \sum_i
\varepsilon_i \varepsilon_i^\ast$, where the sum goes over all
CFHTLenS galaxies in our redshift range.  Therefore, the covariance
between the 184 Clone lines of sight gives us the total covariance
$\mat D + \mat M + \mat V$. Contrary to the case of the 2PCFs
(previous section), this covariance stems from a pure ML estimate, and
therefore the inverse needs to be de-biased by the Anderson-Hartlap
factor $\alpha$. With a typical number of angular scales of $p = 10$
to $15$ the corresponding $\alpha$ is of order $0.9$. We show that our
cosmological results are independent of the number of realisations in
Sect.~\ref{sec:sys}.  Note that for the all derived estimators, the
cosmology-dependence of the covariance is neglected.

For upcoming and future tomographic surveys such as
KiDS\footnote{\texttt{kids.strw.leidenuniv.nl}},
DES\footnote{\texttt{www.darkenergysurvey.org}},
HSC\footnote{\texttt{http://www.naoj.org/Projects/HSC/HSCProject.html}},
Euclid\footnote{\texttt{www.euclid-ec.org}} \citep{2011arXiv1110.3193L} or
LSST\footnote{\texttt{http://www.lsst.org/lsst}}, a much larger suite
of simulations will be necessary. The number of realisations $n$ has
to be substantially larger than the number of bins $p$
\citep{HSS07}. For a multi-bin tomographic shear survey, $p$ can
easily be of the order of several hundreds or more if other probes are
jointly measured such as galaxy clustering or magnification. This
necessitates on the order of a thousand and more independent lines of
sight. This number has to be multiplied by many if a proper treatment
of the cosmology-dependence is to be taken into account. Moreover, a
simple up-scaling of smaller simulated fields to full survey size might
not be easy because of the different area-scaling of the HSV term.

\subsection{Ellipticity calibration corrections}
\label{sec:calib}

We apply the shear calibration as described in \cite{CFHTLenS-sys}, which accounts
for a potential additive shear bias $c$ and multiplicative bias $m$,
\begin{equation}
  \ve^{\rm obs} = (1 + m) \, \ve^{\rm true} + c.
\end{equation}
The additive bias is found to be consistent with zero for
$\varepsilon_1$. The second ellipticity component $\varepsilon_2$
shows a signal-to-noise ratio ($S/N$) and size-dependent bias
which we subtract for each galaxy. This represents a correction which
is on average at the level of $2 \times 10^{-3}$.
The multiplicative bias $m$ is modelled as a function of the galaxy
$S/N$ and size $r$. It is fit simultaneously in 20 bins of $S/N$
and $r$, see \cite{CFHTLenS-shapes}. We use the best-fitting function
$m(S/N, r)$ and perform the global correction to the shear 2PCFs, see
eqs.~(19) and (20) of \cite{CFHTLenS-shapes}.  Accordingly, we
calculate the calibration factor $1 + K$ as the weighted correlation
function of $1 + m$,
\begin{equation}
  1 + K(\vartheta) = \frac{\sum_{ij} w_i w_j (1 + m_i) (1 + m_j)}
  {\sum_{ij} w_i w_j}.
\label{Kestim}
\end{equation}
The final calibrated 2PCFs are obtained by dividing $\xi_+$ and
$\xi_-$ by $1 + K$. The amplitude of $1 + K$ is around 0.91 on all
scales.  The errors on the correlation function from the fit
uncertainty are negligible compared to our statistical
errors. Furthermore, we calculate the covariance matrix $\mat C_m$ for
the correlation function from this uncertainty, and show in
Sect.~\ref{sec:sys} that the cosmological results remain unchanged by
adding this term to the analysis.

\begin{figure}
  \resizebox{\hsize}{!}{
    \includegraphics[bb=50 50 390 302]{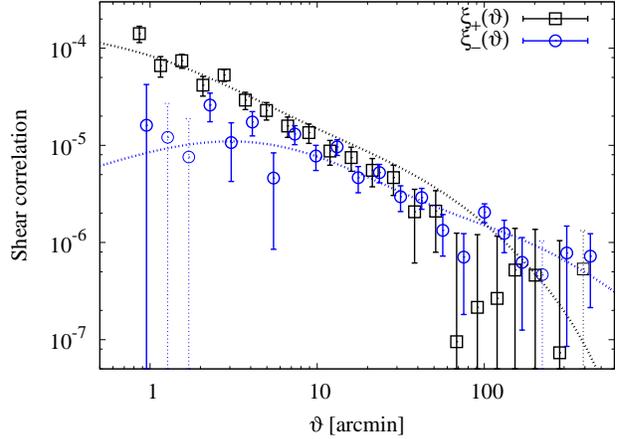}
  }
  
  \caption{The measured shear correlation functions $\xi_+$ (black
    squares) and $\xi_-$ (blue circles), combined from all four Wide
    patches. The error bars correspond to the total covariance
    diagonal. Negative values are shown as thin points with dotted
    error bars. The lines are the theoretical prediction using the
    WMAP7 best-fitting cosmology and the non-linear model described in
    Sect.~\ref{sec:set-up}. The data points and error bars are listed
    in Table \ref{tab:2pcf}.}
  \label{fig:xi-pm.combined}
\end{figure}

\begin{figure}
  \resizebox{\hsize}{!}{
    \includegraphics{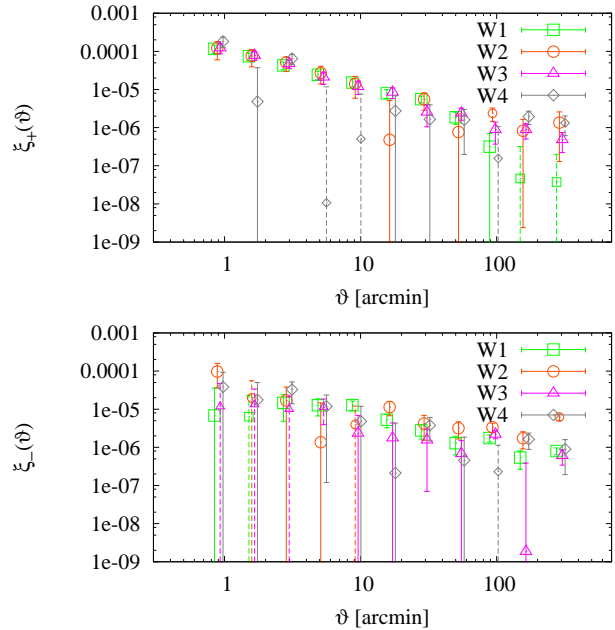}
  }

  \caption{The measured shear correlation functions $\xi_+$
    (\emph{top panel}) and $\xi_-$ (\emph{bottom}), for the four Wide
    patches. The error bars correspond to Poisson noise.
  }
  \label{fig:xi-W1234}
\end{figure}

Figure \ref{fig:xi-pm.combined} shows the combined and corrected 2PCFs,
which are the weighted averages over the four Wide patches with the
number of pairs as weights.
Note that the data points are strongly correlated, in particular $\xi_+$ on scales
larger than about 10 arcmin.
Cosmological results using this data will be presented in Sect.~\ref{sec:results}.
The correlation signal split up into the contributions from the four
Wide patches is plotted in Fig.~\ref{fig:xi-W1234}. There is no
apparent outlier field. The scatter is larger than suggested by the
Poisson noise on large scales, in agreement with the expected cosmic variance.

\subsection{E- and B-modes}
\label{sec:results-EB}

\begin{figure}

    \resizebox{0.72\hsize}{!}{
      \includegraphics[bb=120 -246 390 658]{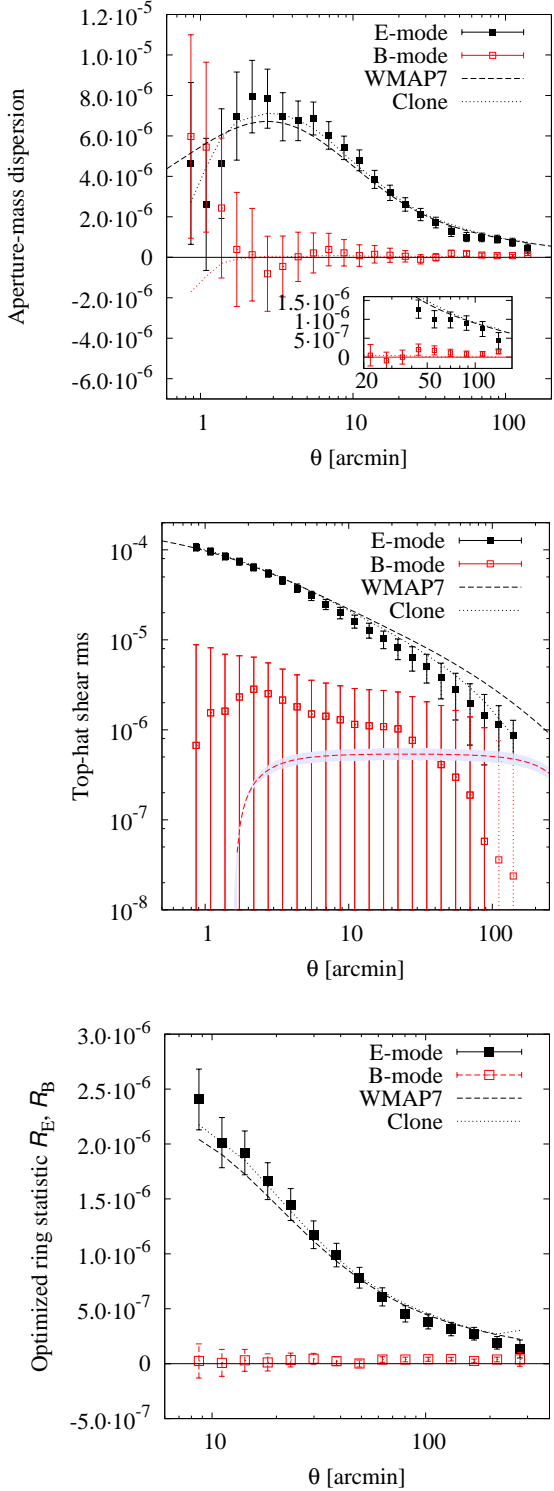}
    }

  \caption{Smoothed second-order functions: Aperture-mass dispersion
    $\langle M_{\rm ap}^2 \rangle$ (\emph{left panel}), shear top-hat
    rms $\langle | \gamma |^2 \rangle$ (\emph{middle}) and optimized
    ring statistic ${\cal R}_{\rm E}$ (\emph{right}), split into the
    E-mode (black filled squares) and B-mode (red open squares).  The
    error bars are the Clone field-to-field rms. The dashed line is
    the theoretical prediction for a WMAP7-cosmology (with zero
    E-/B-mode leakage), the dotted curve shows the Clone
    lines-of-sight mean E-mode signal. For $\langle M_{\rm ap}^2
    \rangle$ and $\langle | \gamma |^2 \rangle$ the WMAP7-prediction
    of the leaked B-mode is shown as red dashed curve; the shaded
    region in the middle panel corresponds to the 95 per cent WMAP7
    confidence interval of $\sigma_8$ (flat $\Lambda$CDM). For the
    shear top-hat rms, negative points are plotted with dashed error
    bars.}
  \label{fig:smoothed}
\end{figure}

The aperture-mass dispersion is shown in the {upper} panel of
Fig.~\ref{fig:smoothed}.  The B-mode is consistent with zero on all
scales. We quantify this by performing a null $\chi^2$ test, taking
into account the B-mode Poisson covariance $\mat C_\times$ as measured
on the Clone,
\begin{equation}
  \chi^2_{\rm B} = \sum_{ij} \left\langle M_\times
  \right\rangle(\theta_i) \left[\mat C_\times^{-1}\right]_{ij} 
  \left\langle M_\times \right\rangle (\theta_j) .
  \label{chi2_null_B}
\end{equation}
Since here the covariance is entirely estimated from the Clone
line-of-sight, the inverse has to be de-biased using the Anderson-Hartlap
factor.  We consider the B-mode over the angular range $[5.5;
140]$ arcmin. As discussed before, the lower scale is where the B-mode due to
leakage is down to a few per cent. The upper limit is given by the
largest scale accessible to the Clone, which is much smaller than the
largest CFHTLenS scale: It is 280 arcmin,
resulting in an upper limit of $\langle M_{\rm ap}^2 \rangle$ of half that
scale.  The resulting $\chi^2 / \rm{dof}$ of $14.9 / 15 = 0.99$ ,
corresponding to a non-null B-mode probability of 46 per cent. Even if we only take
the highest six (positive) data points, we find the $\chi^2$ per
degree of freedom (dof) to be $\chi^2 / \rm{dof} =
4.12 / 6 = 0.69$, which is less than $1\sigma$ significance. The
non-zero B-mode signal at around 50 - 120 arcmin from F08 is not
detected here. 

The top-hat shear rms B-mode is consistent with zero on all measured
scales, as shown in the middle panel of Fig.~\ref{fig:smoothed}. Note,
however, that of all second-order functions discussed in this work,
$\langle|\gamma|^2\rangle$ is the one with the highest correlation
between data points. The predicted leakage from the B- to the E-mode
is smaller than the measured E-mode, but becomes comparable to the latter
for $\theta > 100$ arcmin, where the leakage reaches up to 50 per cent
of the E-mode.

The optimized ring statistic for $\eta = \vt_{\rm min} / \vt_{\rm max}
= 1/50$ is plotted in the {lower} panel of Fig.~\ref{fig:smoothed}. Each
data point shows the E- and B-mode on the angular range between
$\vt_{\rm min}$ and $\vt_{\rm max}$, the latter of which is labelled
on the $x$-axis.  The B-mode is found to be consistent with zero, a
$\chi^2$ null test yields a 35 per cent probability of a non-zero B-mode.

We first test our calculation of COSEBIs on the CFHTLenS Clone with
noise, where we measure a B-mode of at most a few $\times 10^{-12}$
for $n \leq 5$ and $\vt_{\rm max} \leq 250$ arcmin. Even though this
is a few orders of magnitudes larger than the B-mode due to numerical
errors from the estimation from theory, it is insignificant compared
to the E-mode signal.  When including the largest available scales for
the Clone however, $\vt_{\rm max} \sim 280$ arcmin, the B-mode
increases to be of the order of the E-mode. This is true independent
of the binning or whether noise is added. We presume that this is due
to insufficient accuracy with which the shear correlation function is
estimated from the simulation on these very large scales, from only a
small number of galaxy pairs.  Further, for $n > 5$ a similarly large
B-mode is found for some cases of $(\vt_{\rm min}, \vt_{\rm
  max})$. Again, the accuracy of the simulations is not sufficient to
allow for precise numerical integration over the rapidly-oscillating
filter functions of Log-COSEBIs for higher modes
\citep{2012arXiv1208.0068B}. We will therefore restrict ourselves to
$n \leq 5$ for the subsequent cosmological analysis.

\begin{figure}
 
  \resizebox{\hsize}{!}{
    \includegraphics[bb=160 149 440 653]{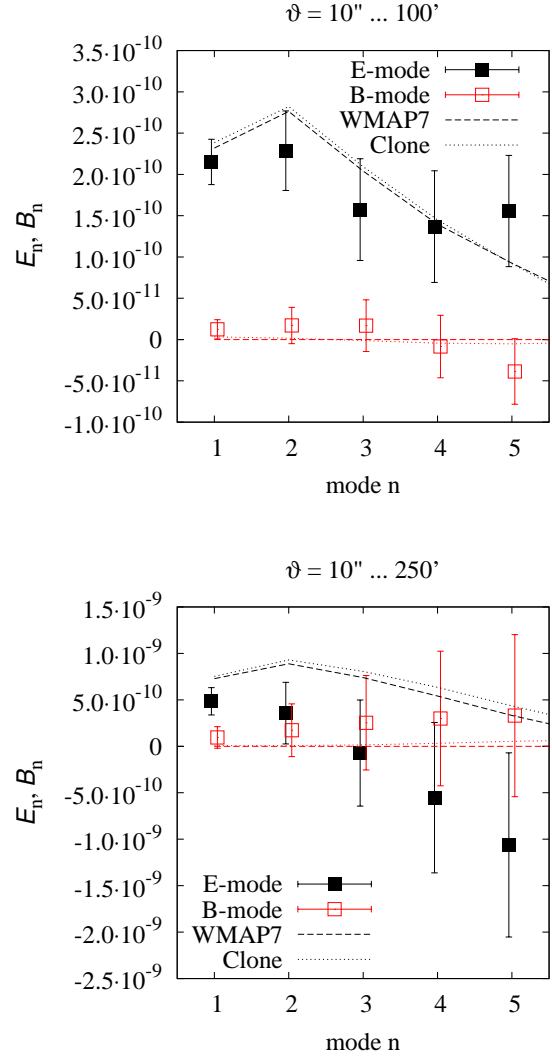}
  }
  
  \caption{COSEBIs with logarithmic filter functions. The left (right)
    panel correspond to a maximum angular scale of $100$ arcmin
    ($250$ arcmin). The filled (open) squares correspond to the
    CFHTLenS E-mode (B-mode). The error bars are the Clone
    field-to-field rms (rescaled to the CFHTLenS area). The dashed
    line is the theoretical prediction for a WMAP7-cosmology, the
    dotted curve shows the Clone mean COSEBIs.}
  \label{fig:cosebis}
\end{figure}

The measured COSEBIs modes are shown in Fig.~\ref{fig:cosebis}.
We use as smallest scale $\vt = 10$ arcsec, and two cases of $\vt_{\rm
  max}$ of 100 and 250 arcmin. In both cases we do not see a
significant B-mode. The signal-to-noise ratio of the high mode points
decreases when the angular range is increased: For $\vt_{\rm max} =
250$ arcmin only the first two modes are significant. This is not
unexpected, since the filter functions for $\vt_{\rm max} = 250$
arcmin sample larger angular scales and put less weight on small
scales where the signal-to-noise ratio in the 2PCFs is larger.


A further derived second-order quantity are the shear E-/B-mode
correlation functions $\xi_{\rm E, B}$ \citep{2002ApJ...568...20C,
  2002ApJ...567...31P}, which have been used in F08.  Whereas they share
the inconvenience with the top-hat shear rms of a formal upper
infinite integration limit, they offer no advantage over the latter, and
will therefore not be used in this work.

\subsection{Conclusion on estimators}
\label{sec:conc_estim}

We compared various second-order real-space shear functions, starting
with the fundamental two-point correlation functions $\xi_{\pm}$. From
the 2PCFs we calculated a number of E-/B-mode separating
functions. The top-hat shear rms $\langle |\gamma|^2 \rangle$ is of
limited use for cosmological analysis because of the
cosmology-dependent E-/B-mode leakage. For the aperture-mass
dispersion $\langle M_{\rm ap}^2 \rangle$ this leakage is confined to
small scales, whereas the optimized ring statistic ${\cal R}_{\rm E}$
and COSEBIs were introduced to avoid any leakage. The
drawback of the 2PCFs is that they are sensitive to large scales outside
the survey area and thus may contain an undetectable B-mode signal
\citep{COSEBIs}.
COSEBIs capture the E-/B-mode signals in an optimal way on a
finite angular-scale interval $[\vt_{\rm min}; \vt_{\rm max}]$. The
interpretation of COSEBIs and the matching of modes to angular
scales are not straightforward since the corresponding filter
functions are strongly oscillating.

For lensing alone, we obtain cosmological parameter constraints on
$\Omegam$ and $\sigma_8$ for the different estimators discussed in
this section. The results and comparisons are presented in
Sect.~\ref{sec:om_s8}.

We decided to use the 2PCFs to compute cosmological constraints in
combination with the other probes, for the following reasons.  The
goal of this paper is to explore the largest scales available for
lensing in CFHTLenS. This is only possible with a sufficiently large
signal-to-noise ratio when using the 2PCFs. We note that on these
large scales our systematics tests, the star-galaxy shape correlation
\citep{CFHTLenS-sys} and the E-/B-mode decomposition (this work), were
not possible. However, since both tests have revealed no systematics
on smaller scales, we are confident that the shear signal up to very
large scales is not significantly contaminated.  Moreover, the
implementation of a cosmology-dependent covariance is currently only
feasible for the 2PCFs.



\section{Cosmology set-up}
\label{sec:setup}

\subsection{Data sets}
\label{sec:data_sets}

We use the following data sets and priors:

\begin{enumerate}

\item CFHTLenS two-point shear correlation functions and covariance as
  described in Sect.~\ref{sec:cfhtlens_data}. We choose the smallest and
  largest angular bins to be 0.9 and 300 arcmin, respectively. This
  includes galaxy pairs between 0.8 and 350 arcmin.

\item Cosmic microwave background (CMB) anisotropies: WMAP7
  \citep{2011ApJS..192...16L, 2010arXiv1001.4538K}. The released WMAP
  code\footnote{\texttt{http://lambda.gsfc.nasa.gov}} is employed to calculate
  the likelihood, see also \cite{WMAP5-Dunkley08}. We use
  \textsc{CAMB}\footnote{\texttt{http://camb.info}} \citep{Lewis:1999bs} to get the
  theoretical predictions of CMB temperature and polarisation power-
  and cross-spectra.

\item Baryonic acoustic oscillations (BAO): SDSS-III (BOSS). We use
  the ratio $D_V/r_{\rm s} = 13.67 \pm 0.22$ of the apparent BAO at
  $z=0.57$ to the sound horizon distance, as a Gaussian random variable,
  from \cite{2012arXiv1203.6594A}.

\item Hubble constant. We add a Gaussian prior for the Hubble constant
  of $h = 0.742 \pm 0.036$ from Cepheids and nearby type-Ia supernovae
  distances from HST \citep[][hereafter R09]{2009ApJ...699..539R}.

\end{enumerate}

In contrast to \cite{KB09} we do not include supernovae of type (SNIa)
Ia. BOSS puts a tight constraint on the expansion history of the
Universe, which is in excellent agreement with corresponding
constraints using the luminosity distance from the most recent
compilation of SNIa \citep{SNLS3, 2012ApJ...746...85S}. Both BOSS and
SNIa are geometrical probes, and adding SNLS to WMAP7+BOSS yields
little improvement on cosmological parameter constraints with the
exception of $w$ \citep{2012MNRAS.425..415S}.

All data sets are treated as independent, neglecting any covariance
between those probes. Experiments which observe the same area on the
sky are certainly correlated since they probe the same cosmological
volume. However, this is a second-order
effect, like CMB lensing, the integrated Sachs-Wolfe effect (ISW), or
the lensing of the baryonic peak \citep{2007PhRvD..75j3509V}. Compared
to the statistical errors of current probes, these correlations can
safely be ignored at present, but have to be taken into account for future
surveys \citep{2012MNRAS.422.2854G}.

\subsection{Sampling the posterior}
\label{sec:sampling}

To obtain constraints on cosmological parameters, we estimate the
posterior density $\pi(\vec p | \vec d, M)$ of a set of parameters
$\vec p$, given the data $\vec d$ and a model $M$. Bayes' theorem
links the posterior to the likelihood $L(\vec d | \vec p, M)$,
the prior $P(\vec p | M)$ and the evidence $E(\vec d | M)$,
\begin{equation}
  \pi(\vec p | \vec d, M) = \frac{L(\vec d | \vec p, M)
    P(\vec p | M)}{E(\vec d | M)}.
\end{equation}
To estimate the true, unknown likelihood distribution $L$, a suite of
$N$-body simulations would be necessary
\citep[e.g.][]{2009A&A...504..689H, 2009A&A...505..969P}.  This is not
feasible in a high-dimensional parameter space, and for the number of
cosmological models probed in this work. Instead, to make progress, we
use a Gaussian likelihood function $L$, despite the fact that neither
the shear field nor the second-order shear functions are Gaussian
random fields. Nevertheless, this is a reasonable approximation, in
particular when CMB is added to lensing \citep{2010PhRvL.105y1301S}.

The construction towards the true likelihood function can be informed
by further features of the estimators, for example constrained
correlation functions \citep{2011A&A...534A..76K}. These constraints
are equivalent to the fact that the power spectrum is positive. We do
however not attempt to make use of these constaints. The
expected deviations are minor compared to the statistical uncertainty
of the data.
The likelihood function is thus given as
\begin{align}
L(\vec d | \vec p, M) = & \, (2 \pi)^{-m/2} |\mat C(\vec p, M)|^{-1/2} \nonumber \\
& \!\!\!\!\!\!\!\! \times \exp\left[ \left( \vec d - \vec
    y(\vec p, M) \right)^{\rm t} \mat{C}^{-1}(\vec p, M) \left( \vec d - \vec
    y(\vec p, M) \right) \right],
\end{align}
where $\vec y(\vec p, M)$ denotes the theoretical prediction for the
data $\vec d$ for a given $m$-dimensional parameter vector $\vec p$ and model $M$.

We sample the posterior with Population Monte Carlo \citep[PMC;][]{WK09,
  KWR10}, using the publicly available code
\textsc{cosmo\_pmc}\footnote{\texttt{www.cosmopmc.info}} \citep{CosmoPMC11}.
PMC is an adaptive importance-sampling technique \citep{CGMR03,
  cappe:douc:guillin:marin:robert:2007} in which
samples $\vec p_n, n = 1 \ldots N$ are created under an
importance function, or proposal density $q$. The sample can be used
as an estimator of the posterior density $\pi$, if each point is weighted
by the normalised importance weight
\begin{equation}
  \bar w_n \propto \frac{\pi(\vec p_n)}{q(\vec p_n)}; \;\;\;
  \sum_{n=1}^N \bar w_n = 1.
\end{equation}
The main difficulty for importance sampling is to find a suitable
importance function. PMC remedies this problem by creating an
iterative series of functions $q_t, t = 1 \ldots T$. In each subsequent
iteration, the importance function is a better representation of the
posterior, so the distribution of importance weights gets progressively
narrower. A measure for this quality of the importance sample is
the normalised Shannon information criterion,
\begin{equation}
  H_{\rm N} = - \sum_{n=1}^N \bar w_n \log \bar w_n.
\end{equation}
As a stopping criterion for the PMC iterations, we use the related
\emph{perplexity} $p$,
\begin{equation}
  p = \exp(H_{\rm N}) / N,
\end{equation}
which lies between 0 and 1, where 1 corresponds to maximum agreement
between importance function and posterior.

Most PMC runs reach values of $p > 0.7$ after 10 or 15
iterations. To obtain a larger final sample, we either perform a last
importance run with five times the number of points, sampled under the
final importance function, or we combine the PMC
samples with the five highest values of $p$. In each iteration we
created 10k sample points; the final sample therefore has 50k points.

An estimate $\hat E$ of the Bayesian evidence
\begin{equation}
  E = \int {\rm d}^m p \, L(\vec d | \vec p, M) P(\vec p | M)
\end{equation}
is obtained at no further computing cost from a PMC simulation \cite{KWR10},
\begin{equation}
  \hat E = \frac 1 N \sum_{n=1}^N w_n.
\end{equation}

\subsection{Theoretical models}
\label{sec:set-up}

We compare the measured second-order shear functions to non-linear
models of the large-scale structure, with a prediction of the density
power spectrum from the \texttt{halofit} fitting formulae of
\citet{2003MNRAS.341.1311S}. For dark-energy models, we adopt the
scheme of the \textsc{icosmo}\footnote{\texttt{www.icosmo.org}} code \citep{2008arXiv0810.1285R}, which
uses the open-CDM fitting formula for a model with $w_0 = -1/3$, and
interpolates between this case and $\Lambda$CDM for models with
differing $w_0$.  This scheme was employed in \citet{SHJKS09} who
compared their non-linear power spectrum with
\citet{2006MNRAS.366..547M} and found good agreement in the range $w_0
\in [-1.5; -0.5]$ out to $k$ of a few inverse Mpc in the relevant redshift
range. \citet{2012PhRvD..85j3518V} have shown that for $\Lambda$CDM
the halofit accuracy of 5 to 10 per cent is sufficient for current
surveys.  From hydro-dynamical simulations, baryonic effects have been
quantified. The results depend on the scenario and specific baryonic
processes included in the simulations. The
bias in the power spectrum to $k$ of few inverse Mpc is between 10 and
20 per cent. The resulting bias on cosmological parameters is smaller
than the CFHTLenS statistical errors \citep{2006ApJ...640L.119J,
  2008ApJ...672...19R, 2011MNRAS.417.2020S}.

We also see a good agreement with the $\Lambda$CDM simulations of
\citet{CFHTLenS-Clone}.  A more accurate non-linear power spectrum on
a wider range of cosmological parameters could be obtained from the Coyote
emulator\footnote{\texttt{http://www.lanl.gov/projects/cosmology/CosmicEmu}}
\citep{CoyoteI, CoyoteII, CoyoteIII}. Unfortunately it is limited in
wave mode ($k < 2.4$ Mpc) and, more importantly, by an upper redshift
of $z = 1$. Due to the scatter in photometric redshifts, we would
have to cut at a very low redshift to limit the redshift tail at $z >
1$. For example, with $z_{\rm ph} \le 0.8$, the fraction of galaxies
at $z > 1$ is 5 per cent, and ignoring these galaxies would bias low the
mean redshift by 0.05 which would result in a bias on $\sigma_8$ which
is larger than the uncertainty in the halofit prescription.
Alternatively, the hybrid approach of \citet{2011MNRAS.tmp.1490E}
could be taken, which pastes the \texttt{halofit} power spectrum to
the Coyote emulator outside the $k$- and $z$-validity range of the
latter. However, this implies multiplying one of the spectra by a
constant to make the combined power spectrum continuous, We do not
deem this sufficiently justified, since this multiplicative factor
does not stem from a fit to numerical simulations and might introduce
a bias to $\sigma_8$.



We individually run PMC for CFHTLenS and WMAP7, respectively.
For the
combined posterior results, we importance sample the WMAP7
final PMC sample, multiplying each sample point with the CFHTLenS
posterior probability.

For weak lensing only, the base parameter vector for the flat
$\Lambda$CDM model is $\vec p = (\Omegam, \sigma_8, \Omega_{\rm b},
n_{\rm s}, h)$. It is complemented by $w_0$ and $\Omega_{\rm de}$ for
dark-energy and non-flat models, respectively.  With CMB, we add the
reionisation optical depth $\tau$ and the Sunyaev-Zel'dovich (SZ) template
amplitude $A_{\rm SZ}$ to the parameter vector. Moreover, we use
$\Delta^2_{\cal R}$ as the primary normalisation parameter, and
calculate $\sigma_8$ as a derived parameter. We use flat priors
throughout which, when WMAP7 is added to CFHTLenS, cover the
high-density regions and the tails of the posterior distribution well.

For model comparison, we limit the parameter ranges to physically
well-motivated priors for those parameters which vary between
models. This is important for any interpretation of the Bayesian
evidence, since the evidence directly depends on the prior. The prior
is an inherent part of the model, and we want to compare physically
well-defined models.

Thus, we limit the total matter and dark-energy densities $\Omegam$ and
$\Omega_{\rm de} \in [0; 1]$, setting a lower physical limit, and
creating a symmetrical prior for the curvature $\Omega_K$ of $[-1;
1]$, which is bound from below by the physical limit of an empty
universe. Note that by sampling both $\Omegam$ and $\Omega_{\rm de}$,
the curvature prior is no longer uniform but has triangular shape.

For the model comparison cases we limit $w_0$ to $[-1;
-1/3]$, therefore excluding phantom energy and dark-energy models
which are non-accelerating at the present time. These priors are the same
as for the models that were compared using the Bayesian evidence in
\cite{KWR10}.  The prior ranges for the other parameters are $\Omegab
\in [0; 0.1], \tau \in
[0.04; 0.2]$, $n_{\rm s} \in [0.7; 1.2]$, $10^9 \Delta^2_{\cal R} \in
[1.8; 3.5]$, $h \in [0.4; 1.2]$ and $A_{\rm SZ} \in [0; 2]$. For the
dark-energy model runs for parameter estimation, which are not used
for model comparison, we use a wide prior on $w_0$ which runs between
$-3.5$ and $0.5$.

\section{Cosmological results}
\label{sec:results}

The most interesting constraints from 2D weak lensing alone are
obtained for $\Omegam$ and $\sigma_8$, which we discuss below for the
four cosmologies considered here. Table \ref{tab:sigma} shows
constraints from lensing alone on the combination $\sigma_8
(\Omega_{\rm m} / 0.27)^\alpha$, which is the direction orthogonal to
the $\Omegam$ - $\sigma_8$ degeneracy `banana' {To obtain
  $\alpha$, we fit a power law to the log-posterior values using
  histograms with optimal bin numbers for estimating the posterior
  density \citep{Scott79}.
}
(Fig.~\ref{fig:CFHTLenS+WMAP_LCDM_curvLCDM_Om_s8}).  We also discuss
constraints on $\Omega_\Lambda$ (for cases with free curvature) and
$w_0$ (for $w$CDM models). Table \ref{tab:parameters} shows the
combined constraints from CFHTLenS+WMAP7 and CFHTLenS+WMAP7+BOSS+R09.
The comparison between cosmological models is shown in Table
\ref{tab:evidence} and described in Sect.~\ref{sec:model_comp}.

\begin{table*}

  \caption{Constraints from CFHTLenS orthogonal to the
    $\Omegam$-$\sigma_8$ degeneracy direction, using the 2PCF. The
    errors are 68\% confidence intervals. The
    four columns correspond to the four different models.}
  \label{tab:sigma}

  \begin{tabular}{|l|l|l|l|l}\hline
    \rule[-3mm]{0em}{8mm}Parameter   & flat\;\;$\Lambda$CDM & flat\;\;$w$CDM	    & curved\;\;$\Lambda$CDM & curved\;\;$w$CDM    \\\hline\hline
    $\sigma_8\;(\Omegam/0.27)^\alpha$ & $0.79\pm0.03$        & $0.79^{+0.07}_{-0.06}$  & $0.80^{+0.05}_{-0.07}$   & $0.82^{+0.05}_{-0.07}$ \\
    $\alpha$                         & $0.59\pm0.02$        & $0.59\pm0.03$	    & $0.61\pm0.02$          & $0.61\pm0.03$       \\ \hline
  \end{tabular}

\end{table*}

\begin{table*}

  \caption{Cosmological parameter results with 68\% confidence
    intervals. The first line for each
    parameters shows CFHTLenS+WMAP7, the second line is
    CFHTLenS+WMAP7+BOSS+R09.
    The star ($^*$) indicates a deduced
    parameter. The four columns correspond to the four
    different models.}
  \label{tab:parameters}

  \renewcommand{\arraystretch}{1}
\begin{tabular}{|l|l|l|l|l|}\hline\hline
\rule[-3mm]{0em}{8mm}Parameter	 &flat\;\;$\Lambda$CDM	 &flat\;\;$w$CDM	 &curved\;\;$\Lambda$CDM	 &curved\;\;$w$CDM	\\\hline\hline
\	 &$0.274^{+0.013}_{-0.012}$	 &$0.325^{+0.082}_{-0.076}$	 &$0.275^{+0.023}_{-0.021}$	 &$0.377^{+0.098}_{-0.079}$	\\
\raisebox{1.5ex}[-1.5ex]{$\Omega_{\textrm{m}}$}	 &$0.283^{+0.010}_{-0.009}$	 &$0.287^{+0.026}_{-0.023}$	 &$0.286^{+0.011}_{-0.010}$	 &$0.271^{+0.028}_{-0.025}$	\\\hline
\	 &$0.815^{+0.016}_{-0.014}$	 &$0.77^{+0.11}_{-0.07}$	 &$0.815^{+0.030}_{-0.025}$	 &$0.715^{+0.090}_{-0.070}$	\\
\raisebox{1.5ex}[-1.5ex]{${\sigma_8}^*$}	 &$0.814^{+0.015}_{-0.014}$	 &$0.809^{+0.039}_{-0.035}$	 &$0.804\pm{0.018}$	 &$0.826^{+0.037}_{-0.039}$	\\\hline
\	 &\	 &$-0.86^{+0.22}_{-0.32}$	 &\	 &$-0.72^{+0.20}_{-0.24}$	\\
\raisebox{1.5ex}[-1.5ex]{$w_0$}	 &\raisebox{1.5ex}[-1.5ex]{$-1$}	 &$-0.99^{+0.11}_{-0.12}$	 &\raisebox{1.5ex}[-1.5ex]{$-1$}	 &$-1.10^{+0.15}_{-0.16}$	\\\hline
\	 &\	 &\	 &$0.726^{+0.016}_{-0.015}$	 &$0.628^{+0.074}_{-0.094}$	\\
\raisebox{1.5ex}[-1.5ex]{$\Omega_{\textrm{de}}$}	 &\raisebox{1.5ex}[-1.5ex]{$1-\Omega_{\textrm{m}}$}	 &\raisebox{1.5ex}[-1.5ex]{$1-\Omega_{\textrm{m}}$}	 &$0.7186^{+0.099}_{-0.014}$	 &$0.735^{+0.028}_{-0.032}$	\\\hline
\	 &\	 &\	 &$-0.0003\pm0.0086$	 &$-0.005^{+0.011}_{-0.012}$	\\
\raisebox{1.5ex}[-1.5ex]{${\Omega_K}^\ast$}	 &\raisebox{1.5ex}[-1.5ex]{$0$}	 &\raisebox{1.5ex}[-1.5ex]{$0$}	 &$-0.0047^{+0.0045}_{-0.0047}$	 &$-0.0063^{+0.0064}_{-0.0045}$	\\\hline
\	 &$0.702^{+0.014}_{-0.013}$	 &$0.66^{+0.11}_{-0.07}$	 &$0.703^{+0.037}_{-0.033}$	 &$0.605^{+0.082}_{-0.062}$	\\
\raisebox{1.5ex}[-1.5ex]{$h$}	 &$0.693\pm{0.010}$	 &$0.691^{+0.032}_{-0.029}$	 &$0.683\pm{0.014}$	 &$0.702^{+0.032}_{-0.030}$	\\\hline
\	 &$0.0456\pm{0.0012}$	 &$0.054^{+0.014}_{-0.013}$	 &$0.0457^{+0.0045}_{-0.0041}$	 &$0.064^{+0.018}_{-0.014}$	\\
\raisebox{1.5ex}[-1.5ex]{$\Omega_{\textrm{b}}$}	 &$0.0465\pm{0.0010}$	 &$0.0471^{+0.0046}_{-0.0042}$	 &$0.0482^{+0.0020}_{-0.0019}$	 &$0.0457^{+0.0047}_{-0.0042}$	\\\hline
\	 &$0.966\pm{0.013}$	 &$0.966\pm{0.014}$	 &$0.965^{+0.013}_{-0.014}$	 &$0.970^{+0.014}_{-0.013}$	\\
\raisebox{1.5ex}[-1.5ex]{$n_{\textrm{s}}$}	 &$0.961\pm{0.012}$	 &$0.959^{+0.013}_{-0.014}$	 &$0.966\pm{0.013}$	 &$0.964^{+0.013}_{-0.014}$	\\\hline
\	 &$0.089^{+0.016}_{-0.014}$	 &$0.088^{+0.016}_{-0.014}$	 &$0.088^{+0.016}_{-0.014}$	 &$0.088^{+0.017}_{-0.013}$	\\
\raisebox{1.5ex}[-1.5ex]{$\tau$}	 &$0.083^{+0.014}_{-0.013}$	 &$0.084^{+0.015}_{-0.013}$	 &$0.088^{+0.016}_{-0.014}$	 &$0.087^{+0.015}_{-0.014}$	\\\hline
\	 &$2.441^{+0.090}_{-0.084}$	 &$2.433^{+0.095}_{-0.087}$	 &$2.445^{+0.095}_{-0.090}$	 &$2.395^{+0.093}_{-0.095}$	\\
\raisebox{1.5ex}[-1.5ex]{$10^9\Delta^2_R$}	 &$2.457^{+0.088}_{-0.081}$	 &$2.465^{+0.097}_{-0.089}$	 &$2.422^{+0.095}_{-0.088}$	 &$2.425^{+0.094}_{-0.089}$	\\\hline
\end{tabular}

\end{table*}

\begin{table*}

  \caption{Bayesian evidence $E$ and Bayes' factor with respect to $\Lambda$CDM,
    for four different cosmological models. The columns for $w_0$ and $\Omega_K$ show
    the prior range for those two parameters. The data is
    CFHTLenS+WMAP7 (fourth column) and CFHTLenS+WMAP7+BOSS+R09 (last
    column), respectively.}
  \label{tab:evidence}

  \begin{tabular}{l|cc|c|c} \hline
  Name                & $w_0$  & $\Omega_K$ & $\ln B_{01}$& $\ln B_{01}$ \\ \hline
  & & & CFHTLenS+WMAP7 & CFHTLenS+WMAP7 \\
  & & & & +BOSS+R09 \\ \hline
  $\Lambda$CDM        & $-1$         & $0$        & $0$    & $0$ \\
  curved $\Lambda$CDM & $-1$         & $[0; 2]$   & $-3.84$ & $-4.0$ \\
  flat $w$CDM         & $[-1; -1/3]$ & $0$        & $0.42$ & $0.58$ \\
  curved $w$CDM       & $[-1; -1/3]$ & $[0; 2]$   & $-3.19$ & $-4.8$ \\ \hline \hline
\end{tabular}

\end{table*}

\subsection{$\Omegam$ and $\sigma_8$}
\label{sec:om_s8}

\paragraph*{Flat $\Lambda$CDM}

\begin{figure}

  \centerline{flat $\Lambda$CDM}

  \resizebox{\hsize}{!}{
    \includegraphics[bb=0 0 360 310]{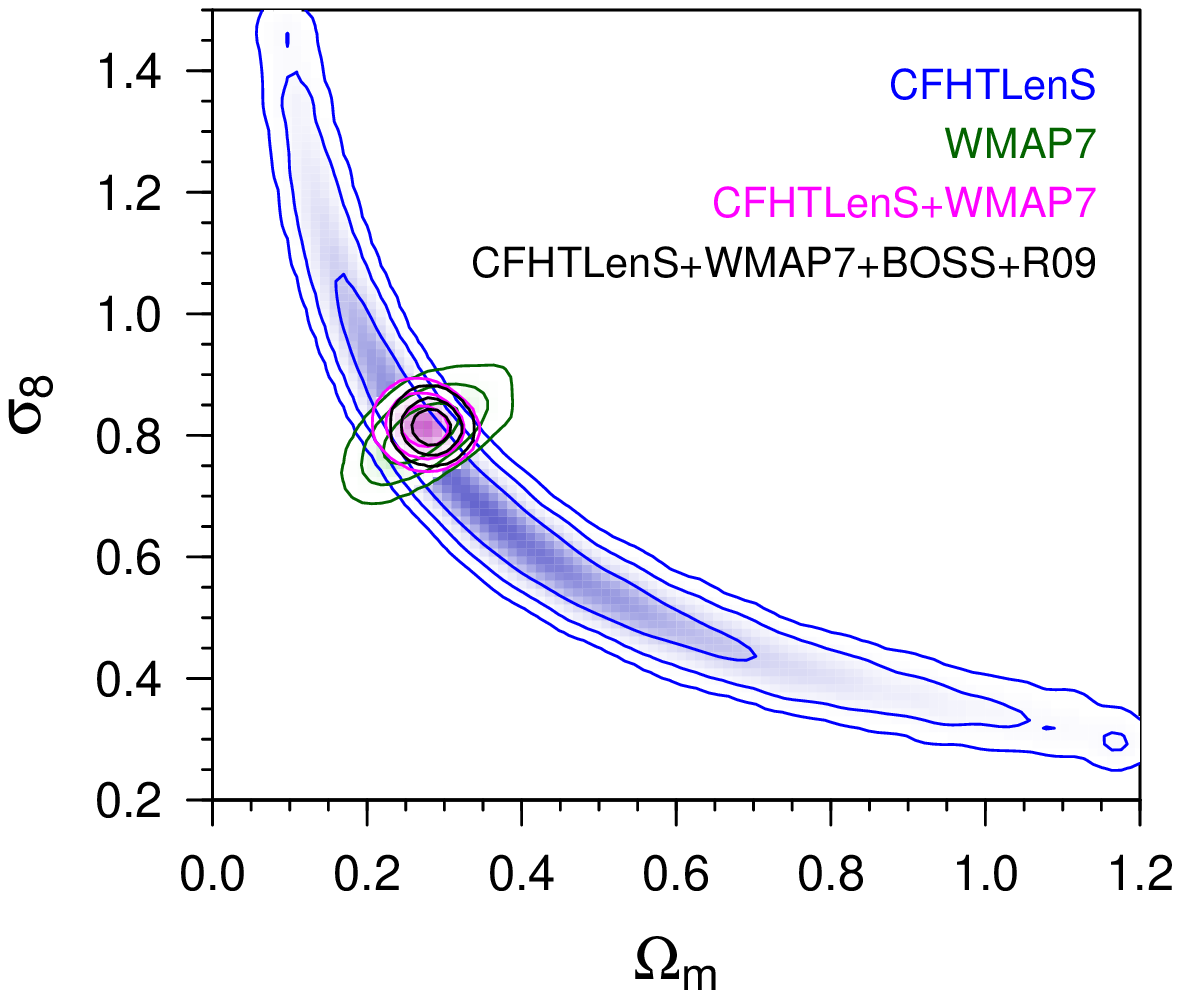}
  }

  \centerline{curved $\Lambda$CDM}%

  \resizebox{\hsize}{!}{
    \includegraphics[bb=0 0 360 310]{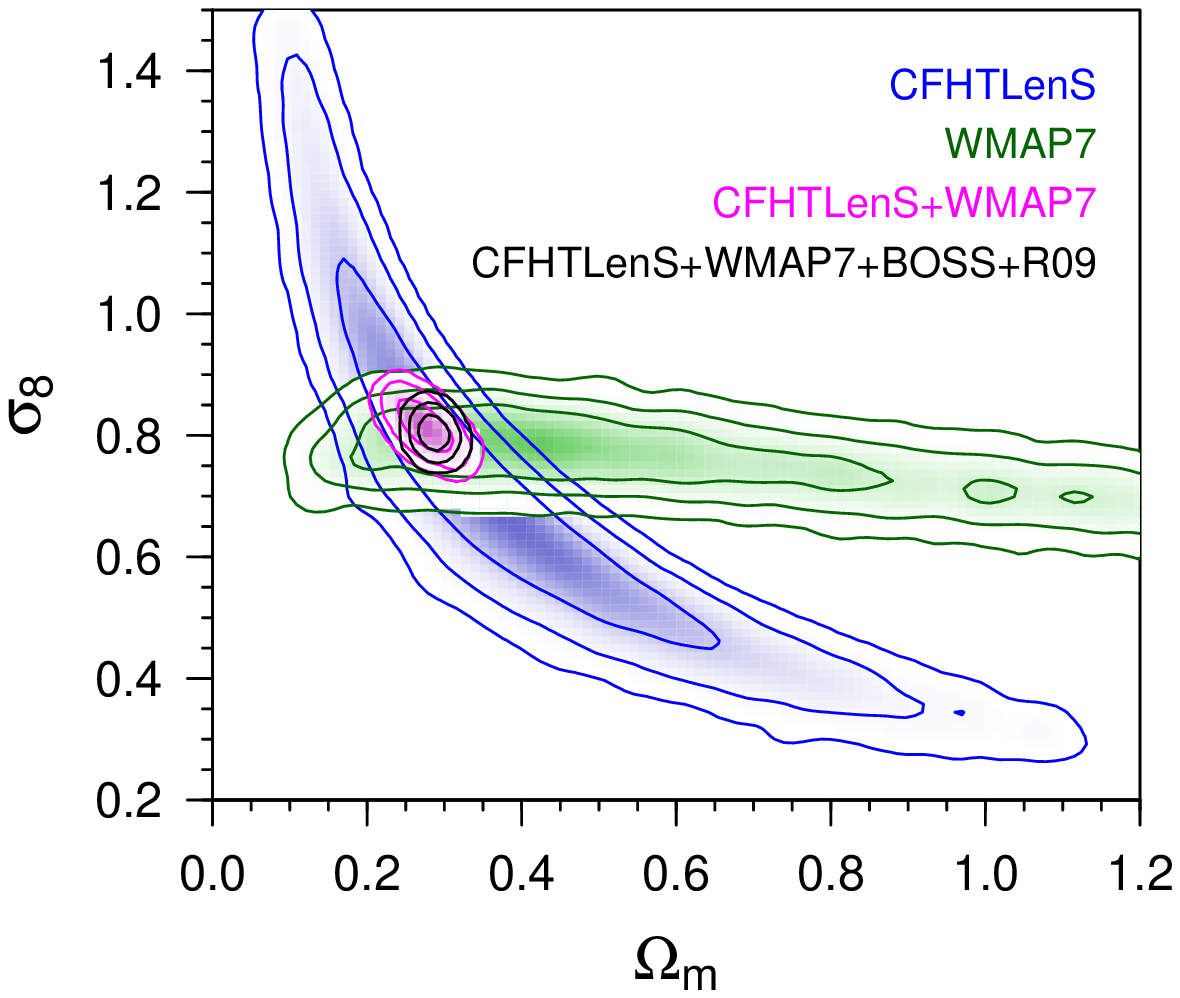}
}

  \centerline{curved $\Lambda$CDM}%
  
  \resizebox{\hsize}{!}{
    \includegraphics[bb=0 0 360 310]{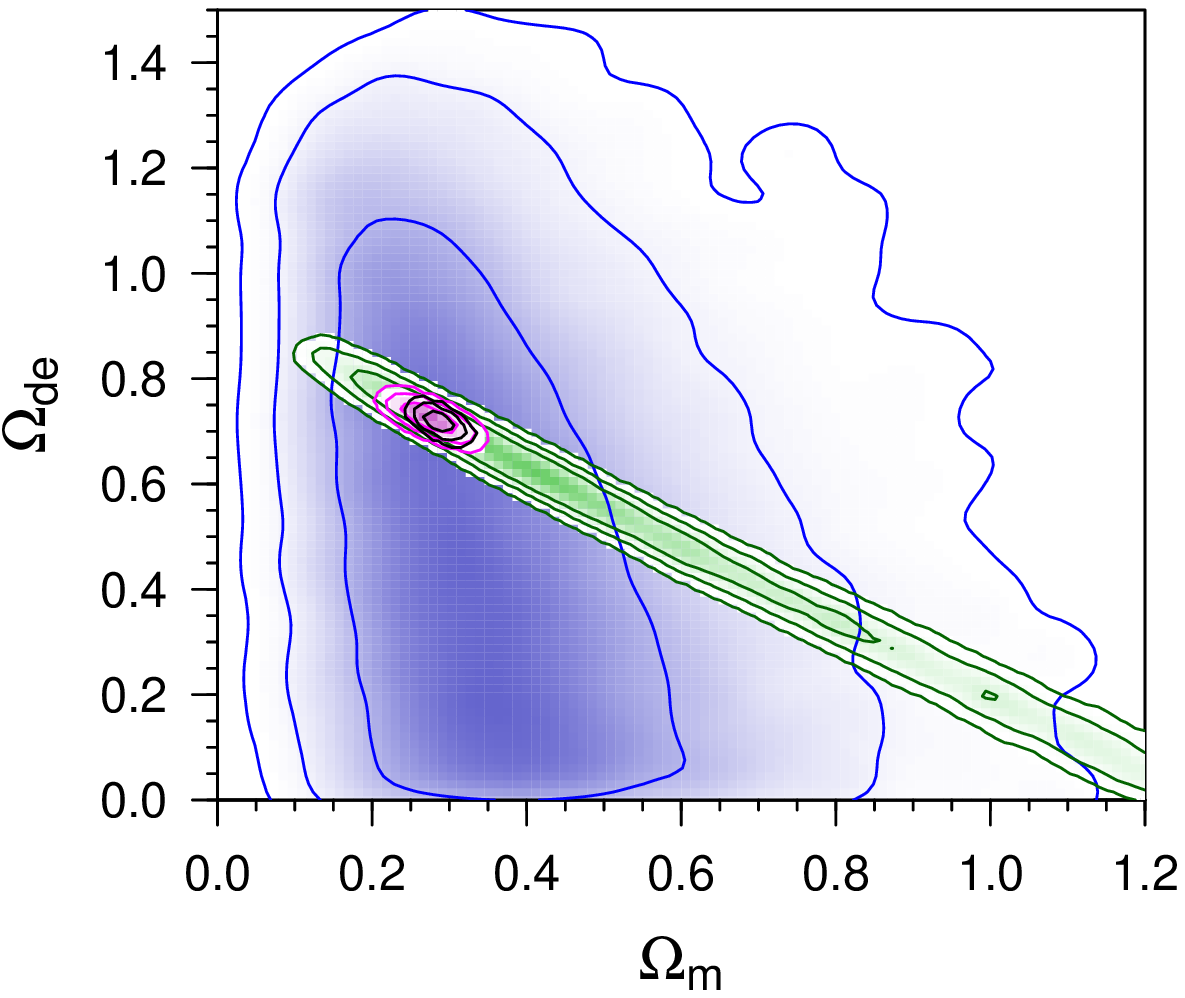}
  }

  \caption{Marginalised posterior density contours (68.3\%, 95.5\%,
    99.7\%) for CFHTLenS (blue contours), WMAP7 (green),
    CFHTLenS+WMAP7 (red) and CFHTLenS+WMAP7+BOSS+R09 (black). The
    model is flat $\Lambda$CDM (\emph{left panel}) and curved
    $\Lambda$CDM (\emph{middle and right panel}), respectively.}

  \label{fig:CFHTLenS+WMAP_LCDM_curvLCDM_Om_s8}
\end{figure}

For a flat $\Lambda$CDM universe, the constraints in the $\Omegam - \sigma_8$ plane 
(left panel of Fig.~\ref{fig:CFHTLenS+WMAP_LCDM_curvLCDM_Om_s8}) 
from CFHTLenS are nearly orthogonal to the ones for WMAP7. CFHTLenS
improves the joint constraints for these parameters by a factor of two.
Lensing plus CMB constrain $\Omegam$ and $\sigma_8$ to better than 5
per cent and 2 per cent, respectively. Adding BOSS and R09 decreases the error on $\Omegam$
to $3.5$ per cent, but does not improve the constraint on $\sigma_8$.

\paragraph*{Flat $w$CDM}

\begin{figure}

  \resizebox{\hsize}{!}{
    \centerline{flat $w$CDM}
  }

  \resizebox{\hsize}{!}{
    \includegraphics[bb=0 0 360 310]{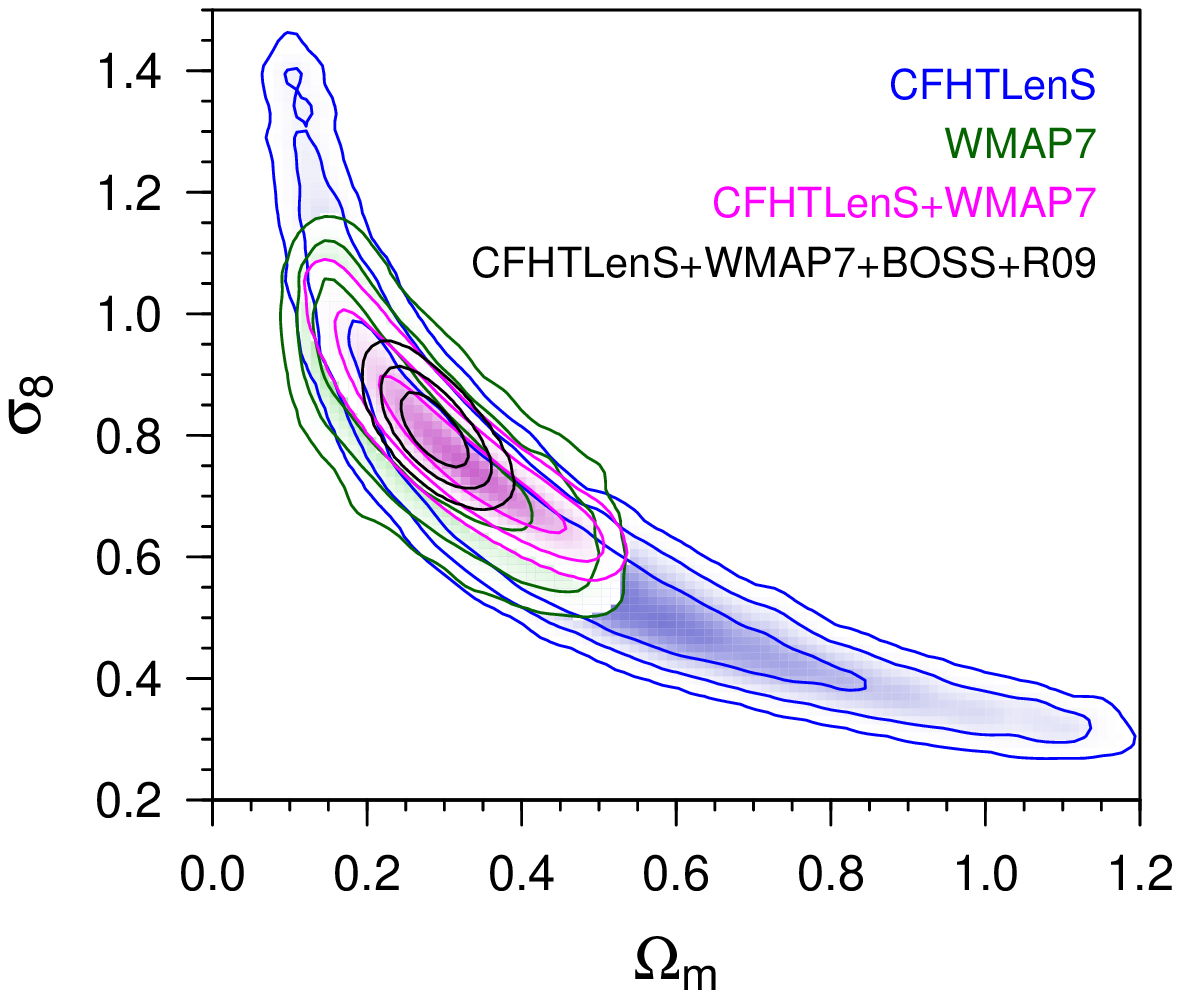}
  }

  \resizebox{\hsize}{!}{
    \centerline{flat $w$CDM}
  }

  \resizebox{\hsize}{!}{
    \includegraphics[bb=0 0 360 310]{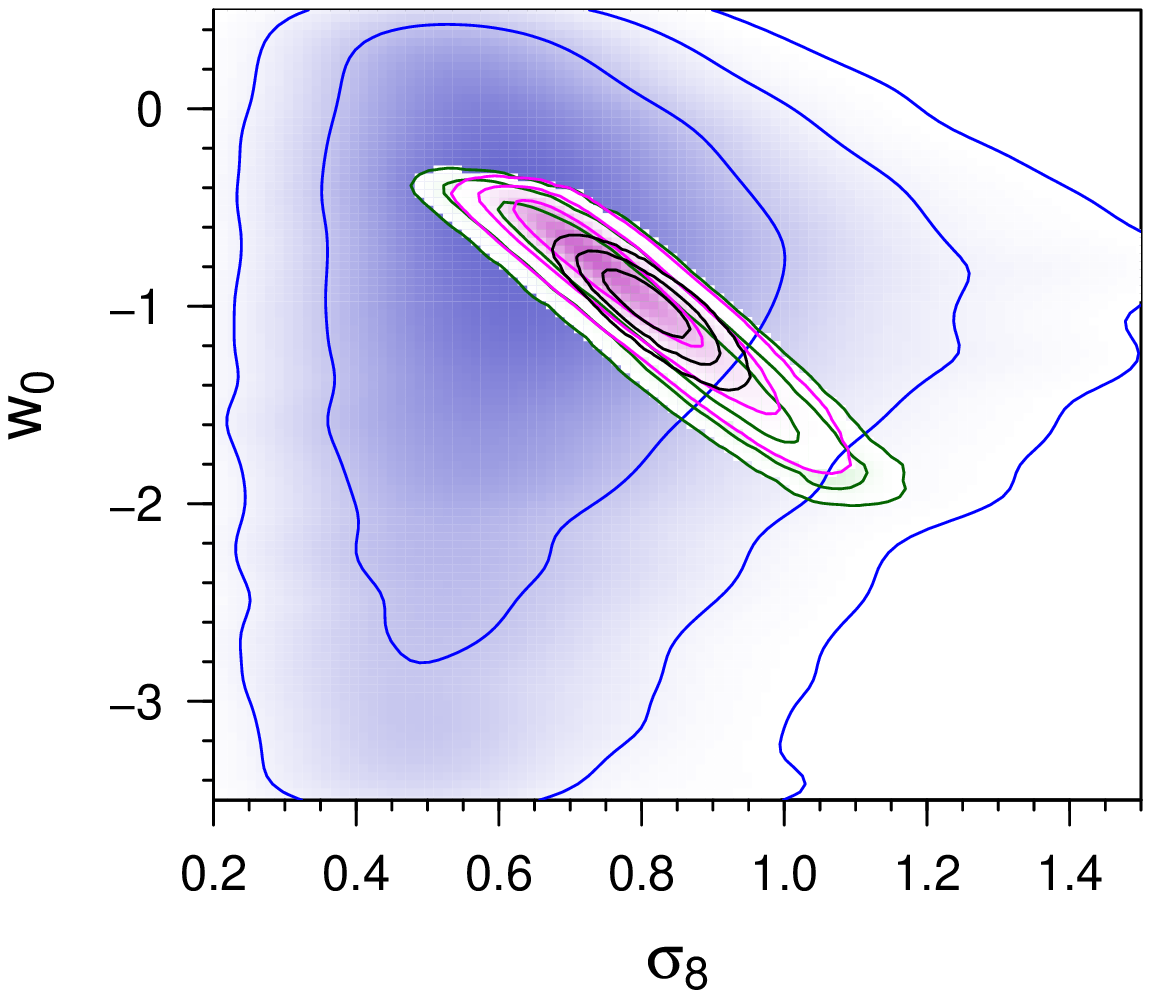}
  }

  \caption{Marginalised posterior density contours (68.3\%, 95.5\%,
    99.7\%) for CFHTLenS (blue contours), WMAP7 (green),
    CFHTLenS+WMAP7 (magenta) and CFHTLenS+WMAP7+BOSS+R09 (black).
    The model is flat $w$CDM.}

  \label{fig:CFHTLenS+WMAP_wCDM}
\end{figure}

If the dark-energy equation-of-state parameter $w_0$ is kept free, CMB
and lensing display the same degeneracy direction between $\Omegam$
and $\sigma_8$ (left panel of Fig.~\ref{fig:CFHTLenS+WMAP_wCDM}). Combining
both probes only partially lifts this degeneracy, the uncertainty on
$\Omegam$ remains at the 25 per cent level. This uncertainty decreases
to 10 percent with the addition of the BOSS BAO distance measure.

The value of the Hubble constant from both CFHTLenS+WMAP7 ($h =
0.66^{+0.11}_{-0.07}$) and BOSS+WMAP7 ($h = 0.65^{+0.08}_{-0.04}$) are
slightly lower when compared to the R09 result, $h = 0.742 \pm 0.036$,
although it is within the 1-$\sigma$ error bar. Since $h$ is degenerate with
all other parameters except $n_{\rm s}$, those
parameter means change with the inclusion of the R09 prior. This
causes the relatively large $\Omegam$ and $\Omegab$ and low
$\sigma_8$ if R09 is not added. The joint Hubble constant with all
four probes is $h = 0.691^{+0.032}_{-0.029}$.

\paragraph*{Curved $\Lambda$CDM}

With curvature left free and no additional priors, CMB anisotropies
cannot determine $\Omegam$ anymore, since there is a degeneracy
between matter density, curvature and the Hubble constant. Lensing
however shows a similar dependency on $\Omegam$ and $\sigma_8$ to the
flat model case. Therefore, the improvement on $\Omegam$ from CFHTLenS
+ WMAP7 with respect to WMAP7 alone is an order of magnitude, to yield
a 8 per cent error. The joint error on $\sigma_8$ is 3.5 per cent.

\paragraph*{Curved $w$CDM}

\begin{figure}

  \resizebox{\hsize}{!}{
    \centerline{curved $w$CDM}
  }

  \resizebox{\hsize}{!}{
    \includegraphics[bb=0 0 360 310]{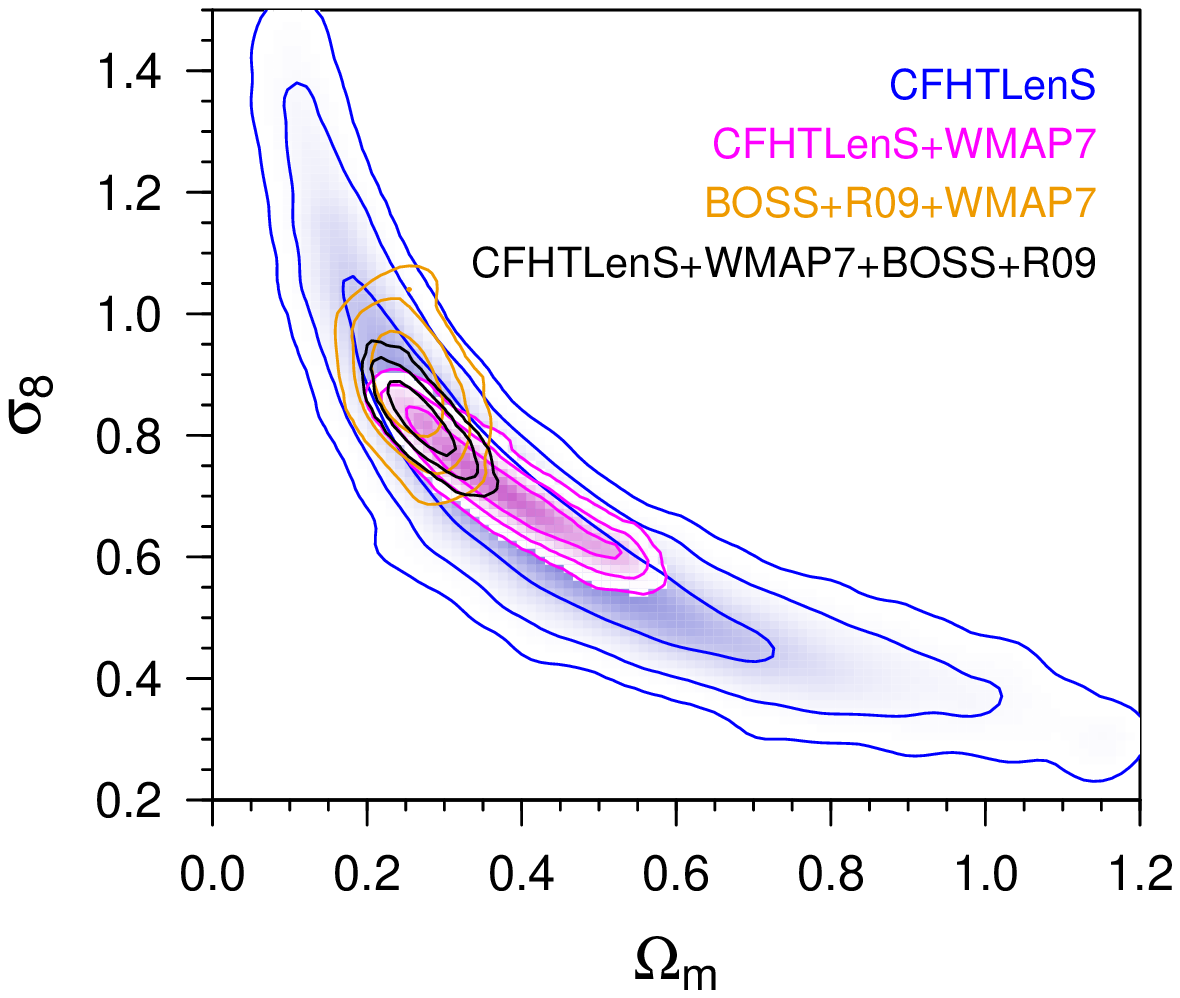}
  }

  \resizebox{\hsize}{!}{
    \centerline{curved $w$CDM}
  }

  \resizebox{\hsize}{!}{
    \includegraphics[bb=0 0 360 310]{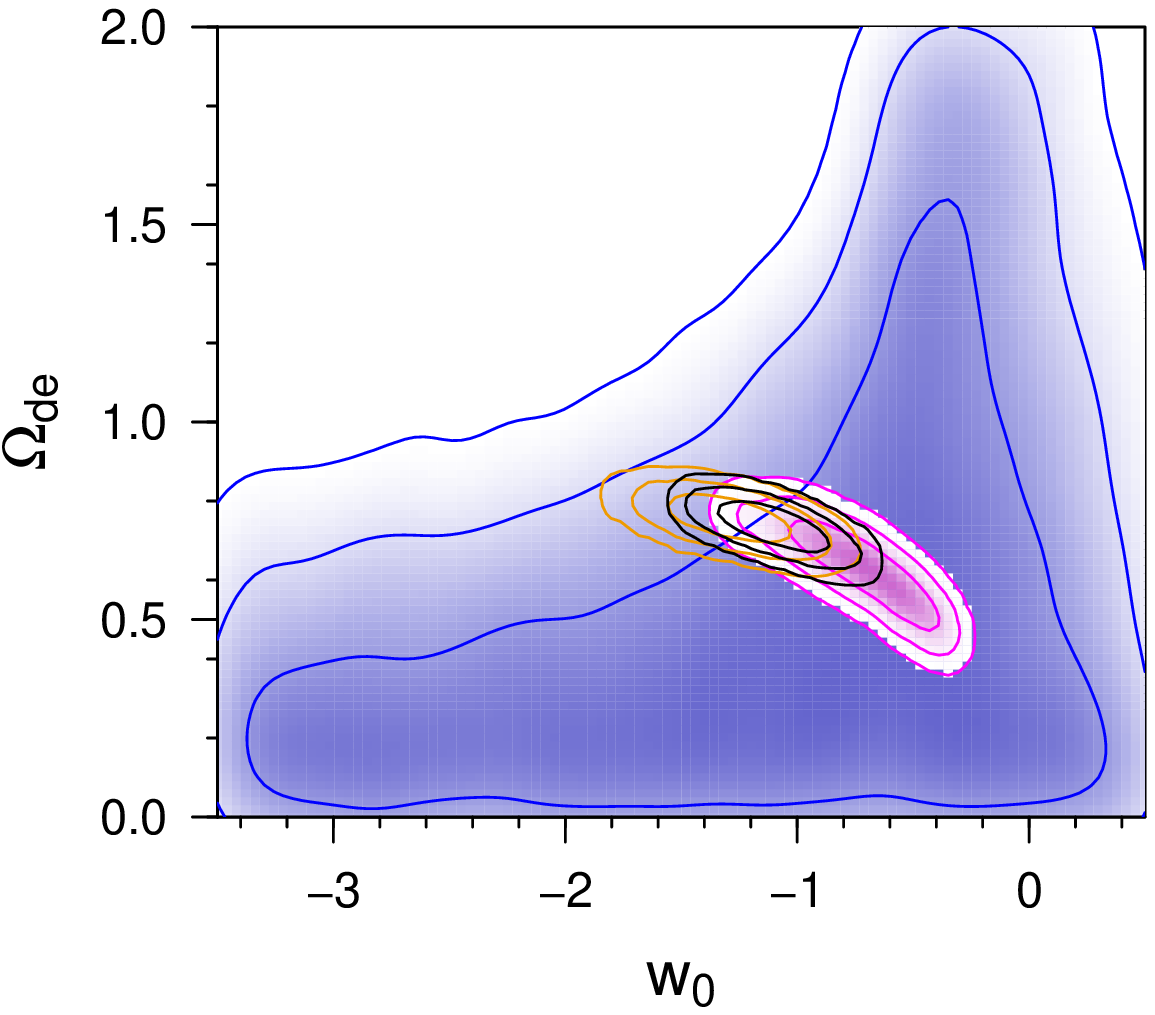}
  }

  \caption{Marginalised posterior density contours (68.3\%, 95.5\%,
    99.7\%) for CFHTLenS (blue contours), WMAP7 (green),
    CFHTLenS+WMAP7 (magenta) and CFHTLenS+WMAP7+BOSS+R09 (black).
    The model is curved $w$CDM.}

  \label{fig:CFHTLenS+WMAP_curvwCDM}
\end{figure}

The $\Omegam$ - $\sigma_8$ degeneracy holds nearly the same as in the
previous cases of models with fewer parameters, as displayed in
the left panel of Fig.~\ref{fig:CFHTLenS+WMAP_curvwCDM}. The value of $\sigma_8
(\Omegam/0.25)^\alpha$ is slightly increased but well within the error
bars. The joint CFHTLenS+WMAP7 results on $\Omegam$ and $\sigma_8$
are similar to the flat $w$CDM case.

The BOSS+R09+WMAP7 results indicate a slightly smaller $\Omegam$ and
larger $\sigma_8$. The joint CFHTLenS+WMAP7+BOSS+R09 allowed region is
therefore on the upper end of the CFHTLenS+WMAP7 banana. The reason
for this is, as in the flat $w$CDM case, the degeneracy of $\Omegam$
and $\sigma_8$ with the Hubble constant. WMAP7 alone prefers a low
value, $h = 0.5^{+0.14}_{-0.13}$, which increases to $h = 0.73 \pm
0.04$ when BOSS+R09 is added. As a consequence, $\Omegam$ decreases
and $\sigma_8$ increases. On the other hand, adding CFHTLenS to WMAP7
leaves the Hubble constant at the relatively low value of $h =
0.60^{+0.08}_{-0.06}$.

\subsection{Dark energy}

For the following results on the dark-energy equation-of-state
parameter $w$, we use the flat prior $[-3.5; 0.5]$.

\paragraph*{Flat $w$CDM}

2D weak gravitational lensing alone is not able to tightly constrain
dark energy, in contrast with 3D tomographic weak lensing. The 68 per
cent confidence limits for $w_0$ (flat $w$CDM) are of order unity,
$w_0 = -1.2^{+0.8}_{-1.4}$. In combination with WMAP7 only, these
errors decrease by a factor of four, and $w_0$ gets constrained to
about 30 per cent.  The CFHTLenS+WMAP7+BOSS constraints on dark energy
are $w_0 = -0.78^{+0.09}_{-0.11}$. We discuss this 
deviation from $\Lambda$CDM in Sect.~\ref{sec:discussion}. Adding the
R09 prior on $H_0$ does not reduce the error but shifts the mean
to the $\Lambda$CDM value, $w_0 = -0.99^{+0.11}_{-0.12}$.

\paragraph*{Curved $w$CDM}

The case of dark energy is similar in the curved case.
CFHTLenS alone results in $w_0 = -1.2^{+0.9}_{-1.8}$.
Adding WMAP7 reduced this uncertainty to 30 per cent.
CFHTLenS+WMAP7+BOSS yield $w_0 = -0.81^{+0.14}_{-0.19}$. Adding the R09 prior on
$H_0$, we find the $\Lambda$CDM-consistent value of  $w_0 = -1.10^{+0.15}_{-0.16}$.

\subsection{Curvature}

CFHTLenS helps to improve the constraint on the curvature density
$\Omega_K$. For $\Lambda$CDM, the uncertainty decreases by a factor of
10 from around $0.1$ (WMAP7 alone) to $0.01$ (CFHTLenS+WMAP7). Adding
BOSS+R09 decreases the error by another factor of two to around
$0.005$. The combined constraints are thus consistent with a flat
universe within $5 \times 10^{-3}$. For a $w$CDM model, this
uncertainty is of the same order.

\begin{figure}

  \resizebox{\hsize}{!}{
    \centerline{flat $\Lambda$CDM}
  }

  \resizebox{\hsize}{!}{
    \includegraphics[bb=0 0 360 310]{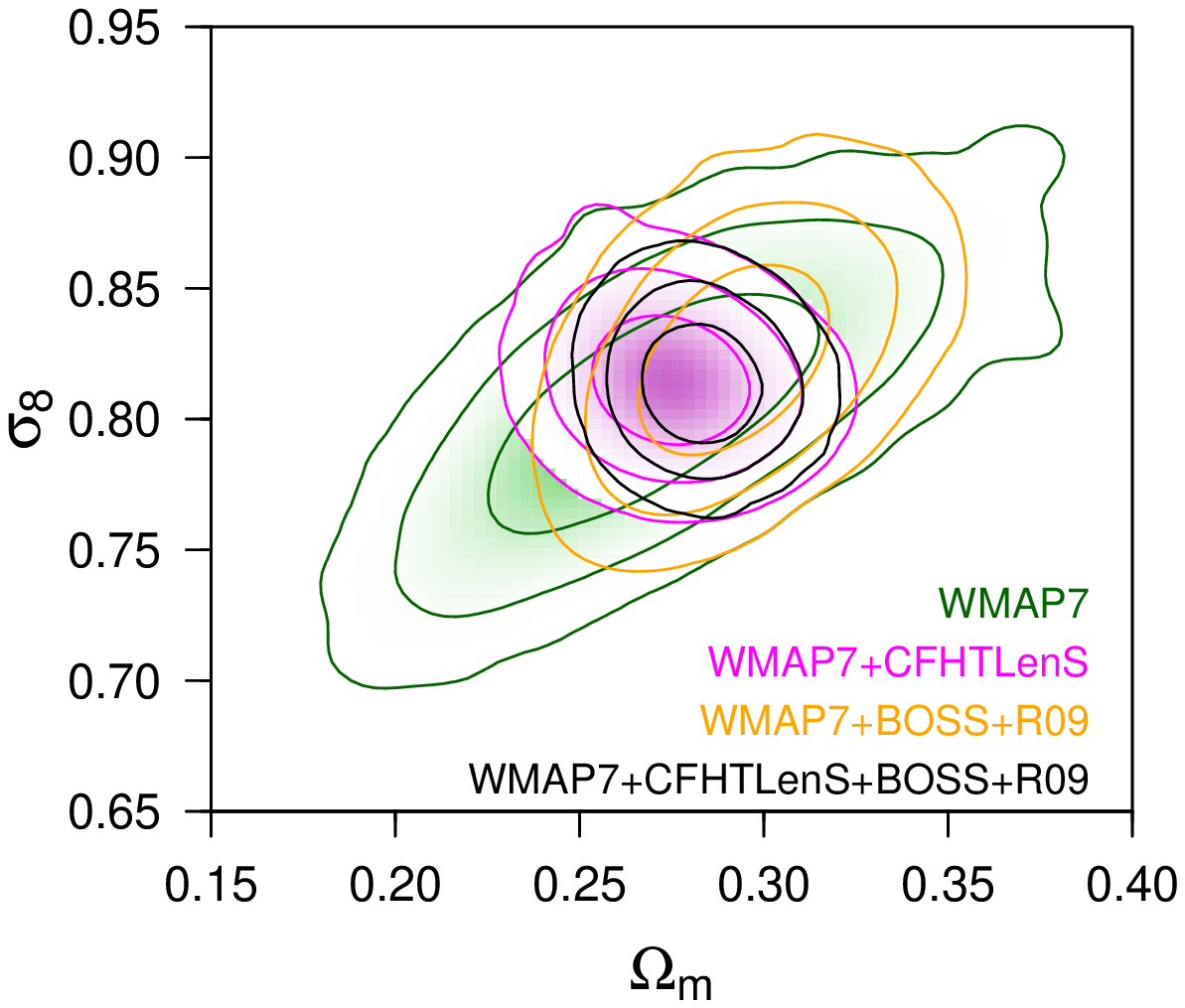}
  }

  \resizebox{\hsize}{!}{
    \centerline{curved $\Lambda$CDM}
  }

  \resizebox{\hsize}{!}{
    \includegraphics[bb=0 0 360 310]{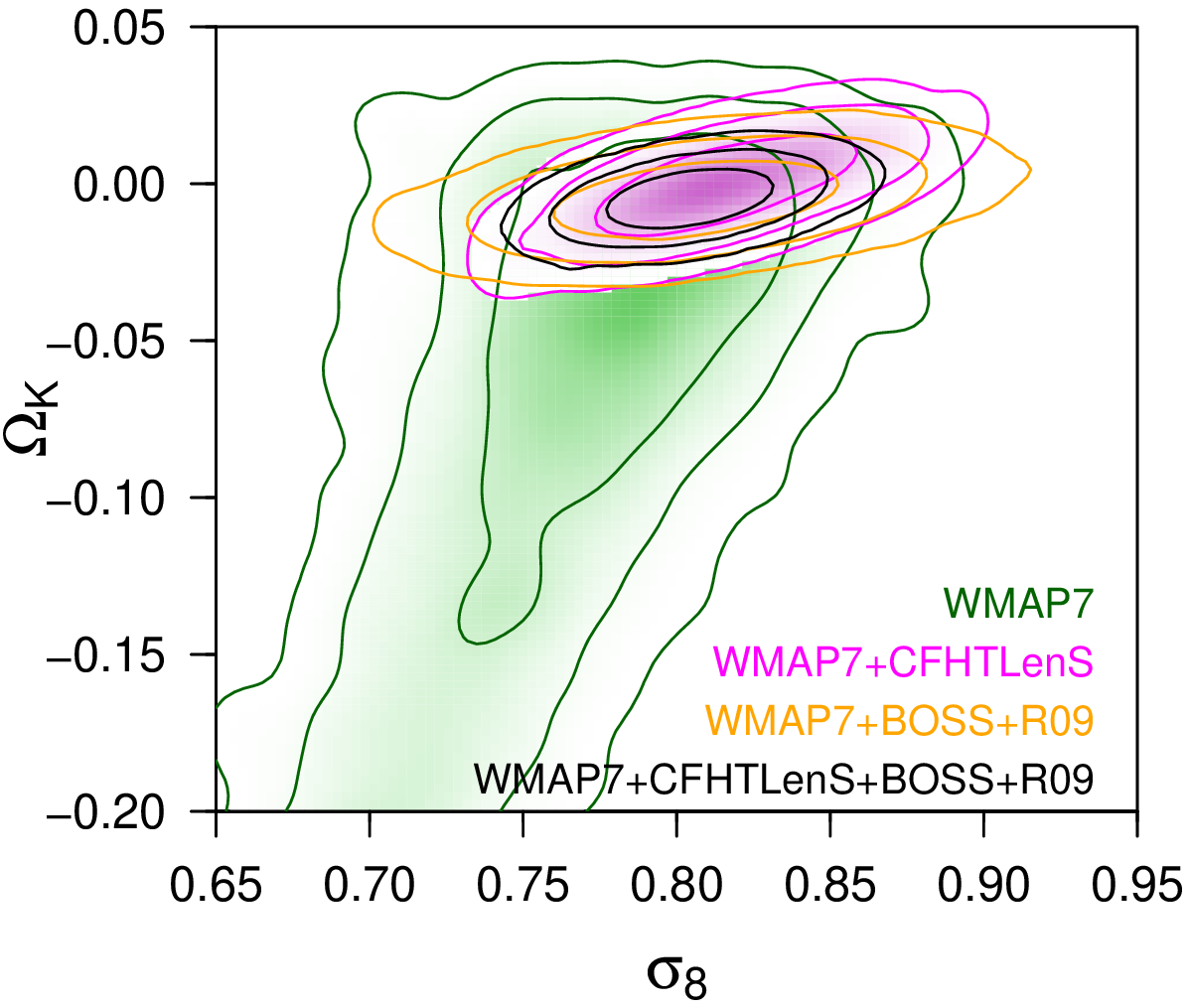}
  }

  \caption{Marginalised posterior density contours (68.3\%, 95.5\%,
    99.7\%) for WMAP7 (green), WMAP7+CFHTLenS (magenta), WMAP7+BOSS+R09 (orange) and
    WMAP7+CFHTLenS+BOSS+R09 (black).  The
    model is flat $\Lambda$CDM (\emph{upper panel}) and curved
    $\Lambda$CDM (\emph{lower panel}), respectively.}

  \label{fig:compare_boss_Lambda}

\end{figure}

\subsection{Model comparison}
\label{sec:model_comp}

In Table \ref{tab:evidence}, the evidence $E$ and the logarithms of
the evidence ratios, $\ln B_{01} = \ln E_0 / E_1$ between the baseline
flat $\Lambda$ model and the other three models are shown. Here, $E_0$
is the evidence for flat $\Lambda$CDM and $E_1$ the evidence for one
of the three models curved $\Lambda$CDM, flat $w$CDM, and curved
$w$CDM, respectively. $B_{01}$ is called the Bayes factor between
model `0' and model `1'.

An empirical scale to interpret such evidence
values was suggested by \citet{jeffreys:1961}, see also
\citet{2008ConPh..49...71T}. Accordingly, two models are not
distinguishable when $| \ln B_{01} | < 1$. If the log-Bayes factor
is between 1 and 2.5, the evidence is called weak. Moderate evidence is
assumed for $2.5 < | \ln B_{01} | < 5$, and strong for values larger than that.

We compute the evidence for the two cases of probes CFHTLenS+WMAP7 and
CFHTLenS+WMAP7+BOSS+R09. Although the evidence for the flat dark
energy model is slightly larger than the one for the cosmological
constant model, the models are indistinguishable: Their respective
evidence values, or posterior odds, are within a factor of two. The
evidence against curved models is moderate, with log-Bayes factor
ratios between $3.2$ and $4.8$, or posterior odd ratios between 25 and
130 in favor of flat $\Lambda$CDM. The significance increases when
adding BOSS and R09, but stays in the moderate range. We remind the
reader that the dark-energy models considered here have the flat prior
$[-1; -1/3]$ for $w$, which corresponds to an accelerating non-phantom
dark energy component.

\section{Comparison of weak-lensing statistics, systematics and
  consistency tests}
\label{sec:tests}

In this section we obtain cosmological constraints from the derived
second-order estimators which were discussed in Sect.~\ref{sec:2pt}.
The following tests are all performed under a
flat $\Lambda$CDM model. The results are listed in Table \ref{tab:sigma_sys}.

\begin{table}

  \caption{Constraints from CFHTLenS orthogonal to the
    $\Omegam$-$\sigma_8$ degeneracy direction. The main result from
    the 2PCF (first row) are compared to other estimators.}

  \label{tab:sigma_sys}

\renewcommand{\arraystretch}{1.5}
\begin{tabular}{|l|l|l|}\hline
\rule[-3mm]{0em}{8mm}Data	               &$\alpha$	 &$\sigma_8\;(\Omegam/0.27)^\alpha$	\\\hline\hline
2PCF	                                       &$0.59\pm0.02$	 &$0.79\pm0.03$	\\
$\langle{M_{\textrm{ap}}^2}\rangle$	       &$0.70\pm0.02$	 &$0.79\pm0.06$	\\
$\langle|\gamma|^2\rangle$ (ignoring offset)   &$0.60\pm0.03$    &$0.78^{+0.04}_{-0.05}$\\
$\langle|\gamma|^2\rangle$ (constant offset)   &$0.58\pm0.03$    &$0.80^{+0.03}_{-0.04}$\\
${\cal{R}}_E$	                               &$0.56\pm0.02$	 &$0.80^{+0.03}_{-0.04}$	\\
COSEBIs ($\vt_{\rm max} = 100\arcmin$)          &$0.60\pm0.02$    &$0.79^{+0.04}_{-0.06}$ \\
COSEBIs ($\vt_{\rm max} = 250\arcmin$)          &$0.64\pm0.03$    &$0.77^{+0.04}_{-0.05}$ \\ \hline
2PCF, constant covariance		       &$0.60\pm0.03$    &$0.78^{+0.03}_{-0.04}$ \\
2PCF ($\vartheta \ge 17\arcmin$)               &$0.65\pm0.02$    &$0.78\pm0.04$         \\
2PCF ($\vartheta \ge 53\arcmin$)      	       &$0.65\pm0.03$    &$0.79^{+0.07}_{-0.06}$  \\   \hline
\end{tabular}
\renewcommand{\arraystretch}{1}

\end{table}

\subsection{Derived second-order functions}
\label{sec:results_other_estim}

As expected, the constraints from the derived second-order estimators
are less tight than from the 2PCFs, since they always involve 
information loss. Moreover, we use a smaller range of
angular scales, cutting off both on the lower and higher end, as
discussed before. All estimators give consistent results.


Aperture-mass dispersion and top-hat shear rms give very similar
constraints compared to the 2PCFs. The position and slope of the
banana are nearly identical, although the width is larger by a factor
two (see Table \ref{tab:sigma}).
For $\langle | \gamma |^2 \rangle$, we analyse two approaches of
dealing with the finite survey-size E-/B-mode leakage:

\begin{enumerate}

\item

  Ignoring the leakage. We fit theoretical models of the top-hat shear
  rms (eqs.~\ref{X_EB_Fourier}, \ref{S}) directly to the measured
  E-mode data points $\langle | \gamma |^2 \rangle(\theta_i)$. Since
  power is lost due to the leakage, we expect $\sigma_8 \, \Omega_{\rm
    m}^\alpha$ to be biased low.

\item 

  We add a constant offset of $5.3 \times 10^{-7}$ to the measured E-mode
  points.  This corresponds to the theoretical leakage for the WMAP7
  best-fitting $\Lambda$CDM model with $\sigma_8 = 0.8$. On scales
  $\theta < 5$ arcmin, the assumption of a constant offset is
  clearly wrong; however, the constant is two orders of magnitudes
  smaller than the measured signal and does ≈not influence the
  result much.

\end{enumerate}

The difference between both cases is about half of the statistical
uncertainty (Table \ref{tab:sigma}).  More sophisticated ways to deal
with this leakage, e.g.~going beyond a constant offset, or
marginalising over a parametrized offset, are expected to yield
similar results. Since they all have the disadvantage of depending on
prior information about a theoretical model which might bias the
result towards that model, we do not consider this second-order estimator further.


The function ${\cal R}_{\rm E}$ on scales between $\vt_{\rm max} = 7.5$ and
$140$ arcmin, for $\eta = \vt_{\rm min} / \vt_{\rm max} = 1/50$
 (implying $\vt_{\rm min} = 9$ arcsec $\ldots 2.8$ arcmin)
yields consistent results to the 2PCFs.
The results for COSEBIs are consistent with the 2PCFs for both cases
of $\vt_{\rm max}$.  For $\vt_{\rm max} = 250$ arcmin we find that the modes which
are consistent with zero do contain information about cosmology. When
only using the first two modes, we obtain $\sigma_8
(\Omegam/0.27)^{0.7} = 0.78^{+0.06}_{-0.07}$, corresponding to a
larger uncertainty of a 30 per cent.

\subsection{Robustness and consistency tests}
\label{sec:sys}

In this section, we test the robustness of our results, by considering
various potential systematic effects, and by varying the angular
scales and estimators.


\paragraph*{Shear calibration covariance}

    We add the shear calibration $\mat C_m$ (Sect.~\ref{sec:calib})
    covariance to the total shear covariance. The correlation
    coefficient of $\mat C_m$ between angular bins is nearly unity, implying that
    the shear calibration varies very little with angular
    separation. Since the magnitude of the covariance is much smaller
    than the statistical uncertainties, the cosmological results are
    virtually unchanged. 

    \paragraph*{Large scales only. }

    The largest ratio of signal-to-noise for cosmic shear is on small,
    non-linear scales. Unfortunately, those scales are the most
    difficult to model, because of uncertainties in the dark-matter
    clustering, and baryonic effects on the total power spectrum. To
    obtain more robust cosmological constraints, we exclude small
    scales from the 2PCFs in two cases, as follows. First, we use the
    cut-off $\vartheta_{\rm c} = 17$ arc minutes. At this scale, the
    non-linear halofit prediction of $\xi_+$ is within 5 per cent of
    the linear model. Baryonic effects, following
    \citet{2011MNRAS.417.2020S}, are reduced to sub per cent
    level. The component $\xi_-$, being more sensitive to small
    scales, is still highly non-linear at this scale. However, since
    most of the constraining power is contained in $\xi_+$, the
    resulting cosmological constraints will not be very sensitive to
    non-linearities. Nevertheless, we use a second, more conservative,
    cut-off of $\vartheta_{\rm c} = 53$ arcmin, where the non-linear
    models of $\xi_-$ is within a factor of two of
    the linear one. On these scales, $\xi_-$ is affected by baryonic
    physics by less than 5 per cent \citep{2011MNRAS.417.2020S}.  In
    both cases, we obtain a mean parameter value for $\sigma_8
    (\Omegam / 0.27)^{0.7}$ which is consistent with the
    result from all angular scales down to an arcmin. In comparison,
    the error bars on this combined parameter are larger by 30 per
    cent for $\vt_{\rm c} = 17$ arcmin, and 100 per cent for $\vt_{\rm
      c} = 53$ arcmin (see Table \ref{tab:sigma}).

    \paragraph*{Reduced shear}

      Since the weak-lensing observable is not the shear $\gamma$, but
      the reduced shear $g = \gamma / (1 - \kappa)$, the relation
      between the shear correlation function and the convergence power
      spectrum ignores higher-order terms \citep[see for an
      overview][]{2009arXiv0910.3786K}. The full calculation of 
      only the third-order terms, involving the convergence
      bispectrum, is very time-consuming and unfeasible for Monte-Carlo
      sampling, requiring the calculation of tens of thousands of
      different models.

      Instead, we explore the fitting formulae from \cite{K10} as a
      good approximation of reduced-shear effects. For a WMAP7
      $\Lambda$CDM cosmology, the ratio between the 2PCFs with and
      without taking into account reduced shear is 1 per cent for $\xi_+$ and
      4 per cent for $\xi_-$ at the smallest scale considered, $\vartheta =
      0.8$ arcmin. Since the fitting formulae are valid within a small
      range around the WMAP7 cosmology, we use them for the combined
      Lensing+CMB parameter constraints. The changes in $\Omegam$
      and $\sigma_8$ for a $\Lambda$CDM model are less than a per cent.

\paragraph*{Number of simulated lines of sight}

        Following \cite{2011arXiv1112.3143H}, we examine the influence
        of the number of simulated lines of sight on the parameter
        constraints.  We calculate the covariance of $\langle M_{\rm
          ap}^2 \rangle$ from 110 instead of 184 lines of sight
        (Sect.~\ref{sec:cov-sm}). Using the corresponding
        Anderson-Hartlap factor $\alpha$, we find identical results as
        before and conclude that the number of simulations is easily
        sufficient for this work.

{
\paragraph*{Non-Gaussian covariance fitting formulae \newline}

	We replace the non-Gaussian covariance of $\xi_+$ calculated from the Clone
	with the fitting formulae from \citet[][S07]{2007MNRAS.375L...6S} and
	\citet[][S11]{2011ApJ...734...76S}, respectively. These works provide prescriptions
	of the non-Gaussian covariance by rescaling the Gaussian cosmic variance term, by fitting
	to $N$-body simulations. Since no recipe for $\xi_-$ is given, we use the Gaussian
	covariance for $\xi_-$ and for the cross-covariance between $\xi_+$ and $\xi_-$.
	We find for $\sigma_8 (\Omegam / 0.27)^\alpha$ the results $0.79^{+0.03}_{-0.04}$ for the
	S07 case, and $0.78^{+0.03}_{-0.04}$ for S11, recovering the mean value with slightly
	larger error bars. This shows that our results are not sensitive to the choice of
	the non-Gaussian covariance.
}

{\paragraph*{Cosmology-independent covariance \newline}
	We compare the two cases of cosmology-dependent covariance (Sect.~\ref{sec:cosmo-dep-cov}),
	and a constant covariance, fixed to the fiducial model. Contrary to \cite{2012arXiv1210.2732J}, we find only
	small differences in the cosmological results. The main effect is a slight increase in the error bars
	for the constant covariance, see Table \ref{tab:sigma}. In particular, around the region of the fiducial
	model, the results are basically the same, and therefore, the joint constraints with other probes are
	virtually unaffected by the choice of the covariance.
}

\section{Discussion and Conclusion}
\label{sec:discussion}

In this paper we present measurements of various second-order shear
correlations from weak gravitational lensing by CFHTLenS, the
Canada-France Hawaii Telescope Lensing Survey. Using a single redshift
bin, $0.2 < z_{\rm p} < 1.3$, we obtain cosmological constraints on
the matter density, $\Omegam$, and the power-spectrum amplitude,
$\sigma_8$. Adding WMAP7, BOSS and R09 data, we obtain parameter
constraints for flat and curved $\Lambda$CDM and dark-energy models,
and calculate the Bayesian evidence to compare the probability for
each model given the data.

\paragraph*{Second-order shear  functions}

Along with the two-point correlation functions $\xi_+$ and $\xi_-$,
which are the fundamental shear observables, we consider various
derived second-order functions, which are able to separate the shear
correlation into its E- and B-mode.  The resulting B-mode is consistent
with zero on all scales. The excess in the E-
and B-mode signal that was seen in the CFHTLS-T0003 data (F08) between
50 and 130 arcmin is no longer present. This excess was most likely
due to systematics in the earlier data, and the removal of this
feature has to be seen as a success of the CFHTLenS
analysis. In particular, hints for deviations from General Relativity
using the F08 data \citep{2010PhRvD..81j3510Z} are not confirmed with
CFHTLenS \citep{CFHTLenS-mod-grav}.

\paragraph*{Cosmological parameters}

The parameter combination which 2D weak lensing can constrain best, is
$\sigma_8 \Omegam^\alpha$ with $\alpha \sim 0.6$. CFHTLenS alone, with
the two-point correlation functions (2PCFs), constrains the
combination $\sigma_8(\Omegam/0.27)^{0.6}$ to $0.787\pm0.032$. To
facilitate a comparison with F08, we write our constraints as
$\sigma_8(\Omegam/0.25)^{\alpha} =  0.82 \pm 0.03$ (2PCFs), $0.84^{+0.03}_{-0.04}$ (top-hat shear rms) and $0.83 \pm 0.06$ (aperture-mass dispersion). The exponent $\alpha$ is around $0.6$ in all three cases.
%
%
%
F08 obtained results with uncertainties between $0.042$ and $0.049$,
although only smoothed second-order quantities were used. The function
closest to the 2PCFs is $\xi_{\rm E}$, for which F08 found 
$\sigma_8 (\Omegam/0.25)^{0.46} = 0.784 \pm 0.049$, corresponding to
an
uncertainty of 6.2 per cent, compared to 4 per cent in this
work. This increase in precision of 50 per cent is
consistent with the naive expectation of CFHTLenS to yield smaller
error bars by a factor of the square root of the area between CFHTLenS
and T0003, which is $\sqrt{129/57} = 1.4$.

Our uncertainty for the top-hat shear rms is about the same as in F08,
whereas the aperture-mass dispersion is slightly more poorly
constrained in this work. Tighter constraints of 5.5 per cent were
found by F08 from $\langle M_{\rm ap}^2 \rangle$, which was used for
their combined CFHTLS-T0003 + WMAP3 results. The fact that the increase
in precision for those smoothed quantities is smaller than expected is
most {due to the limited parameter range of F08, who used tight priors
on the Hubble parameter,  $h \in [0.6; 0.8]$ and fixed $\Omega_{\rm b}$ and
$n_{\rm s}$.  In particular the narrow range of $h$ resulted in tighter
constraints on $\Omegam$ and $\sigma_8$ in F08.}

\cite{SHJKS09} obtained $\sigma_8 (\Omegam/0.3)^{0.6} = 0.68 \pm 0.11$
from a 2D weak-lensing analysis of the COSMOS data. The relatively
large error bars are dominated by cosmic variance from the very small
survey area of 1.64 square degrees, despite the great depth of the
survey.  A 3D lensing analysis with a large number of redshift bins (5
narrow bins up to $z = 4$ and one broad bin with mean redshift of
$1.5$), decreases the error bar by 20 per cent, $\sigma_8
(\Omegam/0.3)^{0.5} = 0.79 \pm 0.09$. This uncertainty is still larger
by a factor of 3 than our CFHTLenS 2D constraints, $\sigma_8
(\Omegam/0.3)^{0.6} = 0.74 \pm 0.03$.

On 168 square degrees of SDSS\footnote{Sloan Digital Sky Survey;
  \texttt{www.sdss.org}} data in the Stripe 82 equatorial region, out
to a median redshift of 0.52, \citet{2011arXiv1112.3143H} recently
obtained $\sigma_8 (\Omegam / 0.264)^{0.67} = 0.65^{+0.12}_{-0.15}$
using a combination of COSEBIs with $(\vt_{\rm min}, \vt_{\rm max}) =
(1.3, 97.5)$ arcmin and an additional data point of
$\xi_+(38$ arcmin$)$. With CFHTLenS we get
$\sigma_8 (\Omegam / 0.264)^{0.6} = 0.80 \pm 0.03$ for the 2PCFs and
$0.80^{+0.04}_{\mathbf{-0.06}}$ for COSEBIs with $\vt_{\rm max} = 100$ arcmin.

{ Recent results from a 2D analysis of the Deep Lens
  Survey\footnote{\texttt{http://dls.physics.ucdavis.edu}} (DLS)
  yielded the very tight constraints $\Omegam = 0.26 \pm 0.05$ and
  $\sigma_8 = 0.87 \pm 0.07$ \citep{2012arXiv1210.2732J}.  Compared to
  CFHTLenS, DLS has a greater depth with mean redshift of $1.1$ and
  $17$ galaxies per square arcmin, but on the other hand covers with 20 square degrees
  a smaller area. The parameter space sampled by DLS
  is similar to F08, with a tight prior on the Hubble constant and fixed
  baryon density $\Omegab$ and spectral index $n_{\rm s}$.
}

For all models of dark-energy and curvature considered here, the
agreement of $\sigma_8$ and $\Omegam$ from CFHTLenS with WMAP7 is very
good. This remains true when BOSS data on the BAO peak is added.
However, we find values of $w_0$ for CFHTLenS+WMAP+BOSS which are
significantly smaller than $-1$, both for flat and curved $w$CDM
models. The reason for this is the near-degeneracy of the dark-energy
parameter with the Hubble constant. The latter takes the rather low
value of around $0.65 \pm 0.1$, which results in a high value of
$w_0$. Adding the R09 result increases $h$ and thus also increases
$w_0$, yielding values which are consistent with $\Lambda$CDM.

For the flat $\Lambda$CDM model, adding CFHTLenS to WMAP7 strongly
helps reducing error bars on $\Omegam$ and $\sigma_8$. The improvement
is larger than in the case where BOSS+R09 is joined with WMAP7, in
particular for $\sigma_8$. The curved $\Lambda$CDM case sees a
slightly different dependence on $\Omega_K$ between
WMAP7+CFHTLenS and WMAP7+BOSS+R09, resulting in tight constraints when
all four probes are combined. Both cases are emphazised in
Fig.~\ref{fig:compare_boss_Lambda}.  In the $w$CDM case, both CFHTLenS
and BOSS cannot improve significantly the dark-energy constraint with
respect to WMAP7. Only the addition of R09, thereby lifting the
$w_0$-$h$ degeneracy decreases the error on the dark-energy parameter.

Our results are in very good agreement with the measurement presented
in \cite{2012ApJ...751L..30H}, who find $\Omega_{\rm m}= 0.259 \pm
0.045$ and $\sigma_8 = 0.748 \pm 0.035$ for a flat $\Lambda$CDM
model. This method uses low- and high-$z$ peculiar velocity data only
and is therefore complementary and independent of our results.

Recent constraints by \citet{2010MNRAS.406.1759M} from the X-ray ROSAT
All-Sky Survey using the cluster mass function for a flat $w$CDM
universe are $\Omegam = 0.23 \pm 0.04$, $\sigma_8 = 0.82 \pm
0.05$, and $w_0 = -1.01 \pm 0.20$, in agreement with the results
presented here. Their relatively low $\Omegam$ is consistent with our
result of $\Omegam = 0.29 \pm 0.02$. When adding CMB (WMAP5), SNIa,
BAO and the cluster gas fraction to the cluster mass function,
\citet{2010MNRAS.406.1759M} get $\Omegam = 0.27 \pm 0.02$.
From the optical SDSS maxBCG cluster catalogue, \citet{2010ApJ...708..645R}
obtain for a flat $\Lambda$CDM model $\sigma_8(\Omegam/0.25)^{0.41} =
0.83 \pm 0.03$.
In combination with WMAP5, they get $\Omegam = 0.265 \pm 0.016$ and
$\sigma_8 = 0.807 \pm 0.020$, which is again consistent with this work.

\paragraph*{Model comparison}

Using the Bayesian evidence, we computed the posterior odds for
various cosmological models. Starting from the basic $\Lambda$CDM
model, we tested extensions of this model which included curvature
$\Omega_K$ and the dark-energy equation of state parameter $w$.
We find no evidence against the standard flat $\Lambda$CDM model.

The constraints for the larger models with free curvature are
consistent with $\Omega_K = 0$. It is therefore not surprising that
those more general models are not favoured over models with fixed flat
geometry. The larger parameter space from the additional degree of
freedom implies a lower predictive capability of those extended models. A
good model should not only predict (a priori) the correct parameter
range where the result is to be found (a posteriori), but also make a
specific and accurate prediction, in other words, it should have a narrow prior
range compared to the posterior. A lack of predictive capability is
penalised by the Bayesian evidence.

In contrast to the two non-flat models (curved $\Lambda$CDM and curved
$w$CDM), the flat $w$CDM universe is indistinguishable from a flat
model with cosmological constant. This can be understood by looking at
the respective additional parameter constraints beyond $\Lambda$CDM,
that is, $\Omega_{\rm de}$ for the curved and $w_0$ for
$w$CDM. Compared to the corresponding prior, the allowed posterior
range for $\Omega_{\rm de}$ is a lot smaller than the one for $w_0$
since the latter parameter is less tightly constrained. Therefore, the
curved models are less predictive, corresponding to a lower evidence.
Both the very tight constraints on $\Omega_K$, with error of about
$0.005$, as well as the moderate Bayesian evidence in favour of a flat
model strengthen the emerging picture that we live indeed in a
Universe with zero curvature.

\section*{Acknowledgments}

We would like to thank the anonymous referee for useful suggestions
which helped to improve the manuscript. Further, we thank P.~Schneider
for insightful comments and discussions. This work is based on
observations obtained with MegaPrime/MegaCam, a joint project of the
Canada-France-Hawaii Telescope (CFHT) and CEA/Irfu, at CFHT which is
operated by the National Research Council (NRC) of Canada, the
Institut National des Sciences de l'Univers (INSU) at the Centre
National de la Recherche Scientifique (CNRS) of France, and the
University of Hawaii. This research used the facilities of the
Canadian Astronomy Data Centre operated by the NRC of Canada with the
support of the Canadian Space Agency. We thank the CFHT staff, in
particular J.-C.~Cuillandre and E.~Magnier, for the observations, data
processing and continuous improvement of the instrument
calibration. We also thank TERAPIX for quality assessment, and
E.~Bertin for developing some of the software used in this
study. CFHTLenS data processing was made possible thanks to support
from the Natural Sciences and Engineering Research Council of Canada
(NSERC) and HPC specialist O.~Toader. The N-body simulations were
performed on the TCS supercomputer at the SciNet HPC Consortium. The
early stages of the CFHTLenS project were made possible thanks to the
European Commissions Marie Curie Research Training
Network DUEL (MRTN-CT-2006-036133) and its support of CFHTLenS team members LF, HHi, and BR.

MK is supported in parts by the Deutsche Forschungsgemeinschaft (DFG)
cluster of excellence ``Origin and Structure of the Universe''. LF
acknowledges support from NSFC grants 11103012 \& 10878003, Innovation
Program 12ZZ134 and Chen Guang project 10CG46 of SMEC, and STCSM grant
11290706600 \& Pujiang Program 12PJ1406700. CH and FS acknowledges
support from the European Research Council (ERC) through grant
240185. TE is supported by the DFG through project ER 327/3-1 and the
Transregional Collaborative Research Centre TR 33. HHo acknowledges
support from Marie Curie IRG grant 230924, the Netherlands
Organisation for Scientific Research (NWO) through grant 639.042.814
and from the ERC through grant 279396. HHi is supported by the Marie
Curie IOF 252760 and by a CITA National Fellowship. TDK is supported
by a Royal Society University Research Fellowship. YM acknowledges
support from CNRS/INSU and the Programme National Galaxies et
Cosmologie (PNCG). LVW and MJH acknowledge support from NSERC. LVW
also acknowledges support from the Canadian Institute for Advanced
Research (CIfAR, Cosmology and Gravity program). BR acknowledges
support from the ERC through grant 24067, and the Jet Propulsion
Laboratory, California Institute of Technology (NASA). TS acknowledges
support from NSF through grant AST-0444059-001, SAO through grant
GO0-11147A, and NWO.  ES acknowledges support from the NWO grant
639.042.814 and support from ERC under grant 279396. MV acknowledges
support from NWO and from the Beecroft Institute for Particle
Astrophysics and Cosmology.

{\small Author Contribution: All authors contributed to the
  development and writing of this paper. The authorship list reflects
  the lead authors of this paper (MK, LF, CH and FS) followed by two
  alphabetical groups. The first group includes key contributers to
  the science analysis and interpretation in this paper, the founding
  core team and those whose long-term significant effort produced the
  final CFHTLenS data product. The second group covers members of the
  CFHTLenS team who made a significant contribution to either the
  project, this paper or both, and external authors. JHD, SV and LVW
  produced the numerical simulations and, with CH, created the clone.
  The CFHTLenS collaboration was co-led by CH and LVW, and the
  CFHTLenS Cosmology Working Group was led by TDK.  }

\label{lastpage}

\bibliographystyle{mn2e}
\bibliography{astro}

\begin{appendix}

\section{Filter functions}
\label{sec:filter_functions}

We give expressions of the filter functions $F_+, F_-$, needed to
calculate the derived second-order shear observables from the shear
correlation functions (eq.~\ref{X_EB}). See Table \ref{tab:F+-} for
the relation between $F_\pm$ and the following functions and the
integration ranges.

\paragraph*{Aperture-mass dispersion}

The filter functions for the aperture-mass dispersion, defined in
\citet{2002A&A...389..729S}, are for $x<2$
\begin{align}
T_+(x) = & \frac{6(2-15x^2)}{5}\left[1-\frac 2 \pi
  \arcsin(x/2)\right] + \frac{x \sqrt{4-x^2}}{100 \pi} \nonumber  \\
   & \times \left(120+2320 x^2 -
754x^4+132x^6-9x^8 \right) ;  \nonumber\\
T_-(x)  = &  \frac{192}{35 \pi} x^3
\left(1-\frac{x^2}{4}\right)^{7/2},
\label{T}
\end{align}
The functions have finite support, and are set to zero for $x>2$. 
The Fourier-space filter function for the aperture-mass dispersion (eq.~\ref{X_EB_Fourier})
is
\begin{equation}
\hat U(\ell) = \hat U_\theta(\ell) =
\frac{24 {\rm J}_4(\theta \ell)}{(\theta \ell)^2}.
\label{Uhat-map}
\end{equation}

\paragraph*{Top-hat shear rms}

For the top-hat shear rms, the real-space filter functions are
\begin{align}
  S_+(x) = & \frac{1}{\pi} \left [4 \arccos(x/2)  - x\sqrt{4-x^2}\right
  ] {\rm H}(2-x) \ ;  \nonumber\\
  S_-(x) = & \frac{1} {\pi x^4} \times \nonumber\\
  & \left[ { x\sqrt{4-x^2}(6-x^2)-8(3-x^2) \arcsin(x/2)} \right ]\!\!,
  \label{S}
\end{align}
where ${\rm H}$ is the Heaviside step function. Thus, $S_+$ has
support $[0; 2]$ wheras $S_-$ has infinite support.
The Fourier transform of $S_+$ is
\begin{equation}
\hat U(\ell) = \hat U_\theta(\ell) = \frac{2 {\rm J}_1(\theta
  \ell)}{(\theta \ell)^2}.
\label{Uhat-gamma}
\end{equation}

\paragraph*{Optimized ring statistic\newline}

{To obtain an E-/B-mode decomposition of the 2PCF on a finite angular range
$[\vartheta_{\rm min}; \vartheta_{\rm max}]$ via the sum in eq.~\ref{X_EB},
two integral conditions for the filter
function $F_+$ need to be fulfilled \citep{SK07}:
\begin{equation}
\int_{\vt_{\rm min}}^{\vt_{\rm max}} \dd \vt \, \vt F_+(\vt)
  =  \int_{\vt_{\rm min}}^{\vt_{\rm max}} \dd \vt \, \vt^3 F_+(\vt)
  = 0.
  \label{F+cond}
\end{equation}
The function $F_-$ can be obtained by an integral over $F_+$, which follows from the
relation eq.~\ref{X_EB_Fourier}, see \citet{2002A&A...389..729S}. Apart from these
conditions, the functions $F_\pm$ can be freely chosen.

For the optimized ring statistics,} the filter functions corresponding to
${\cal R_{\rm E, B}}$ are linear combinations
of Chebyshev polynomials of the second kind,
\begin{align}
  T_{+}(\vt) = &
\tilde T_+\left(x = \frac{2\vt - \vt_{\rm max} - \vt_{\rm min}}{\vt_{\rm max} - \vt_{\rm min}} \right) 
\nonumber \\
  = & \sum_{n=0}^{N-1} a_n U_n(x); \label{TEB} \\
  U_n(x) = & \frac{\sin[(n+1) \arccos x]}{\sin(\arccos x)}.
\end{align}
The coefficients $a_n$ can be chosen freely, under the condition that
${\cal R}_{\rm E, B}$ are pure E- and B-mode components, respectively. We
take the $a_n$ from \cite{FK10}, which minimized the
$\Omegam$-$\sigma_8$ 1$\sigma$-area using the CFHTLS-T0003 survey
setting, and for fixed $\eta = \vt_{\rm min} / \vt_{\rm max} =
1/50$. For a fixed $\eta$, ${\cal R}_{\rm E, B}$ depends on only one
angular scale $\theta$, which we take to be $\vt_{\rm max}$.

\paragraph*{COSEBIs}

The COSEBIs filter functions we use here are polynomials in the logarithm of the
angular scale $\theta$,
\begin{align}
  T^{\rm log}_{+, n}(\vt) = &
  t^{\rm log}_{+, n}\left[z = \ln \left(\frac{\vt}{\vt_{\rm min}}\right)\right]
\nonumber \\
  = & 
  N_n \sum_{j=0}^{n+1} c_{nj} z^j
  = N_n \prod_{j=1}^{n+1} (z - r_{nj}).  \label{t}
\end{align}
{
The polynomials $t^{\rm log}_{+, n}(z)$ have been constructed in \cite{COSEBIs} using
eq.~(\ref{F+cond}) as
an orthonormal and complete set of functions. The coefficients $c_{nj}$ are fixed
by integral conditions that assure the E-/B-mode decomposition of the 2PCF on a finite
angular integral. They are given by a linear system of equations, which
is given in 
}
%
\cite{COSEBIs}. To solve this system, a very high
numerical accuracy is needed. We use the \textsc{Mathematica} program given
in \cite{COSEBIs} to obtain the coefficients for a given $\vt_{\rm
  min}$ and $\vt_{\rm max}$, and store the zeros $r_{ni}$, for which a lower
accuracy is sufficient. The function $F_-$ is then calculated using
eqs.~(38) and (39) from \cite{COSEBIs}.

Both for COSEBIs and for FK10, no closed expressions for the Hankel
transforms of $T_{\pm}$ have been found (yet); neither for the
Fourier-space counterparts, apart form the linear COSEBIs
\citep{2012A&A...542A.122A}.  To obtain the theoretical predictions for
these functions, we first calculate $\xi_{\pm}$ via eq.~(\ref{xipmestim}),
and use eq.~(\ref{X_EB}) to obtain the COSEBIs prediction.

To calculate the numerical integration over the correlation function
(eq.~\ref{X_EB}) with high enough precision, we split up the interval $[0;
z_{\rm max}$] into 10 pieces, and perform a Romberg-integration on
each piece with relative precision of $10^{-6}$. The resulting
numerical B-mode is smaller than $10^{-15}$ for modes $n \leq 10$,
which is about three orders of magnitudes smaller than the predicted
E-mode.

\section{CFHTLenS second-order weak-lensing data}
\label{sec:data_tables}

The measured 2PCFs as shown in Fig.~\ref{fig:xi-pm.combined}. We list
the data points and the total error in Table \ref{tab:2pcf}. The full
covariance is available on request or via the web page
\texttt{http://cfhtlens.org}. The derived second-order E- and B-mode
functions are listed in Tables \ref{tab:map2} to \ref{tab:cosebis_250}.

\begin{table}

  \caption{The CFHTLenS two-point correlation functions (2PCFs) $\xi_+$
    and $\xi_-$, for different angular scales $\vartheta$, see
    Sect.~\ref{sec:xi}. The values
    of $\sigma$ indicate the error from the total covariance diagonal
    (Sect.~\ref{sec:grafting}). The covariance is the one at the fiducial Clone
    cosmology.}
  \label{tab:2pcf}

  \tabcolsep1ex
  \renewcommand{\arraystretch}{1}
\begin{tabular}{lllll}\hline\hline
\rule[-3mm]{0em}{8mm}$\vartheta/'$	 &$\xi_+(\vartheta)$	 &$\sigma[(\xi_+(\vartheta)]$	 &$\xi_-(\vartheta)$	 &$\sigma[(\xi_-(\vartheta)]$	\\\hline
\rule[-1mm]{0em}{5mm}$0.9$	 &$1.411\cdot10^{-4}$	 &$2.686\cdot10^{-5}$	 &$1.610\cdot10^{-5}$	 &$2.621\cdot10^{-5}$	\\
$1.2$	 &$6.619\cdot10^{-5}$	 &$1.586\cdot10^{-5}$	 &$-1.209\cdot10^{-5}$	 &$1.491\cdot10^{-5}$	\\
$1.6$	 &$7.438\cdot10^{-5}$	 &$1.223\cdot10^{-5}$	 &$-7.580\cdot10^{-6}$	 &$1.123\cdot10^{-5}$	\\
$2.2$	 &$4.162\cdot10^{-5}$	 &$9.507\cdot10^{-6}$	 &$2.600\cdot10^{-5}$	 &$8.486\cdot10^{-6}$	\\
$2.9$	 &$5.298\cdot10^{-5}$	 &$7.438\cdot10^{-6}$	 &$1.067\cdot10^{-5}$	 &$6.426\cdot10^{-6}$	\\
$3.9$	 &$2.923\cdot10^{-5}$	 &$5.864\cdot10^{-6}$	 &$1.738\cdot10^{-5}$	 &$4.892\cdot10^{-6}$	\\
$5.2$	 &$2.287\cdot10^{-5}$	 &$4.669\cdot10^{-6}$	 &$4.607\cdot10^{-6}$	 &$3.755\cdot10^{-6}$	\\
$7.0$	 &$1.583\cdot10^{-5}$	 &$3.745\cdot10^{-6}$	 &$1.306\cdot10^{-5}$	 &$2.892\cdot10^{-6}$	\\
$9.4$	 &$1.351\cdot10^{-5}$	 &$3.045\cdot10^{-6}$	 &$7.760\cdot10^{-6}$	 &$2.255\cdot10^{-6}$	\\
$12.5$	 &$8.737\cdot10^{-6}$	 &$2.494\cdot10^{-6}$	 &$9.643\cdot10^{-6}$	 &$1.770\cdot10^{-6}$	\\
$16.8$	 &$7.487\cdot10^{-6}$	 &$2.088\cdot10^{-6}$	 &$4.652\cdot10^{-6}$	 &$1.402\cdot10^{-6}$	\\
$22.4$	 &$5.536\cdot10^{-6}$	 &$1.791\cdot10^{-6}$	 &$5.241\cdot10^{-6}$	 &$1.126\cdot10^{-6}$	\\
$30.0$	 &$4.656\cdot10^{-6}$	 &$1.605\cdot10^{-6}$	 &$2.959\cdot10^{-6}$	 &$8.791\cdot10^{-7}$	\\
$40.2$	 &$2.072\cdot10^{-6}$	 &$1.457\cdot10^{-6}$	 &$2.901\cdot10^{-6}$	 &$7.040\cdot10^{-7}$	\\
$53.7$	 &$2.104\cdot10^{-6}$	 &$1.310\cdot10^{-6}$	 &$1.332\cdot10^{-6}$	 &$6.072\cdot10^{-7}$	\\
$71.7$	 &$9.524\cdot10^{-8}$	 &$1.153\cdot10^{-6}$	 &$7.075\cdot10^{-7}$	 &$5.260\cdot10^{-7}$	\\
$95.5$	 &$2.149\cdot10^{-7}$	 &$9.903\cdot10^{-7}$	 &$2.048\cdot10^{-6}$	 &$4.496\cdot10^{-7}$	\\
$125.3$	 &$2.660\cdot10^{-7}$	 &$8.990\cdot10^{-7}$	 &$1.240\cdot10^{-6}$	 &$4.552\cdot10^{-7}$	\\
$160.3$	 &$5.207\cdot10^{-7}$	 &$8.782\cdot10^{-7}$	 &$6.247\cdot10^{-7}$	 &$4.994\cdot10^{-7}$	\\
$211.7$	 &$4.607\cdot10^{-7}$	 &$9.043\cdot10^{-7}$	 &$-4.670\cdot10^{-7}$	 &$5.690\cdot10^{-7}$	\\
$296.5$	 &$7.331\cdot10^{-8}$	 &$9.729\cdot10^{-7}$	 &$7.811\cdot10^{-7}$	 &$6.959\cdot10^{-7}$	\\
\hline
\end{tabular}

\end{table}

\begin{table}

  \caption{The CFHTLenS aperture-mass dispersion. The E-mode, $\langle M_{\rm ap}^2 \rangle$, and B-mode $\langle M_\times \rangle$,
    are given for different angular smoothing scales $\theta$, see
    Sect.~\ref{sec:results-EB}. The values
    of $\sigma$ indicate the error from the scaled Clone covariance diagonal
    (Sect.~\ref{sec:cov-sm}). Note that for cosmological results, we do not use scales below 5 arcmin.}
  \label{tab:map2}

  \tabcolsep1ex
  \renewcommand{\arraystretch}{1}
\begin{tabular}{lllll}\hline\hline
\rule[-3mm]{0em}{8mm}$\theta/'$	 &$\langle{M_\textrm{ap}^2}\rangle(\theta)$	 &$\sigma[\langle{M_\textbf{ap}^2}\rangle(\theta)]$	 &$\langle{M_\times^2}\rangle(\theta)$	 &$\sigma[\langle{M_\times^2}\rangle(\theta)]$	\\\hline
\rule[-1mm]{0em}{5mm}$0.9$	 &$4.640\cdot10^{-6}$	 &$3.997\cdot10^{-6}$	 &$5.970\cdot10^{-6}$	 &$3.962\cdot10^{-6}$	\\
$1.1$	 &$2.615\cdot10^{-6}$	 &$3.266\cdot10^{-6}$	 &$5.444\cdot10^{-6}$	 &$3.202\cdot10^{-6}$	\\
$1.4$	 &$4.631\cdot10^{-6}$	 &$2.709\cdot10^{-6}$	 &$2.436\cdot10^{-6}$	 &$2.541\cdot10^{-6}$	\\
$1.7$	 &$6.978\cdot10^{-6}$	 &$2.178\cdot10^{-6}$	 &$3.884\cdot10^{-7}$	 &$2.065\cdot10^{-6}$	\\
$2.2$	 &$7.939\cdot10^{-6}$	 &$1.794\cdot10^{-6}$	 &$1.274\cdot10^{-7}$	 &$1.718\cdot10^{-6}$	\\
$2.7$	 &$7.842\cdot10^{-6}$	 &$1.460\cdot10^{-6}$	 &$-8.143\cdot10^{-7}$	 &$1.424\cdot10^{-6}$	\\
$3.5$	 &$6.946\cdot10^{-6}$	 &$1.191\cdot10^{-6}$	 &$-4.514\cdot10^{-7}$	 &$1.147\cdot10^{-6}$	\\
$4.4$	 &$6.747\cdot10^{-6}$	 &$9.810\cdot10^{-7}$	 &$2.530\cdot10^{-8}$	 &$8.793\cdot10^{-7}$	\\
$5.5$	 &$6.861\cdot10^{-6}$	 &$8.139\cdot10^{-7}$	 &$2.168\cdot10^{-7}$	 &$7.060\cdot10^{-7}$	\\
$6.9$	 &$6.023\cdot10^{-6}$	 &$6.835\cdot10^{-7}$	 &$3.835\cdot10^{-7}$	 &$5.902\cdot10^{-7}$	\\
$8.7$	 &$5.409\cdot10^{-6}$	 &$5.809\cdot10^{-7}$	 &$2.212\cdot10^{-7}$	 &$4.846\cdot10^{-7}$	\\
$11.0$	 &$4.793\cdot10^{-6}$	 &$5.157\cdot10^{-7}$	 &$8.553\cdot10^{-8}$	 &$3.813\cdot10^{-7}$	\\
$13.9$	 &$3.851\cdot10^{-6}$	 &$4.448\cdot10^{-7}$	 &$1.480\cdot10^{-7}$	 &$3.138\cdot10^{-7}$	\\
$17.5$	 &$3.187\cdot10^{-6}$	 &$3.809\cdot10^{-7}$	 &$1.037\cdot10^{-7}$	 &$2.439\cdot10^{-7}$	\\
$22.0$	 &$2.612\cdot10^{-6}$	 &$3.350\cdot10^{-7}$	 &$5.001\cdot10^{-8}$	 &$1.950\cdot10^{-7}$	\\
$27.7$	 &$2.113\cdot10^{-6}$	 &$3.015\cdot10^{-7}$	 &$-9.627\cdot10^{-8}$	 &$1.619\cdot10^{-7}$	\\
$35.0$	 &$1.718\cdot10^{-6}$	 &$2.656\cdot10^{-7}$	 &$-7.630\cdot10^{-10}$	 &$1.374\cdot10^{-7}$	\\
$44.1$	 &$1.269\cdot10^{-6}$	 &$2.410\cdot10^{-7}$	 &$1.976\cdot10^{-7}$	 &$1.160\cdot10^{-7}$	\\
$55.5$	 &$1.002\cdot10^{-6}$	 &$2.279\cdot10^{-7}$	 &$1.783\cdot10^{-7}$	 &$1.040\cdot10^{-7}$	\\
$70.0$	 &$9.834\cdot10^{-7}$	 &$2.109\cdot10^{-7}$	 &$1.116\cdot10^{-7}$	 &$9.397\cdot10^{-8}$	\\
$88.2$	 &$9.004\cdot10^{-7}$	 &$1.920\cdot10^{-7}$	 &$8.229\cdot10^{-8}$	 &$9.364\cdot10^{-8}$	\\
$111.1$	 &$7.437\cdot10^{-7}$	 &$1.985\cdot10^{-7}$	 &$8.539\cdot10^{-8}$	 &$1.028\cdot10^{-7}$	\\
$140.0$	 &$4.320\cdot10^{-7}$	 &$2.181\cdot10^{-7}$	 &$1.412\cdot10^{-7}$	 &$1.540\cdot10^{-7}$	\\
\hline
\end{tabular}

\end{table}

\begin{table}

  \caption{The CFHTLenS optimized ring statistic. The E-mode, $\mathcal{R}_{\rm E}$, and B-mode $\mathcal{R}_{\rm B}$,
    are given for different angular smoothing scales $\theta$, see
    Sect.~\ref{sec:results-EB}. The values
    of $\sigma$ indicate the error from the scaled Clone covariance diagonal
    (Sect.~\ref{sec:cov-sm}).}
  \label{tab:REB}

  \tabcolsep1ex
  \renewcommand{\arraystretch}{1}
\begin{tabular}{lllll}\hline\hline
\rule[-3mm]{0em}{8mm}$\theta/'$	 &$\mathcal{R}_\textrm{E}$	 &$\sigma[\mathcal{R}_\textrm{E}]$	 &$\mathcal{R}_\textrm{B}$	 &$\sigma[\mathcal{R}_\textrm{E}]$	\\\hline
\rule[-1mm]{0em}{5mm}$8.7$	 &$2.405\cdot10^{-6}$	 &$2.768\cdot10^{-7}$	 &$2.457\cdot10^{-8}$	 &$1.555\cdot10^{-7}$	\\
$11.1$	 &$2.012\cdot10^{-6}$	 &$2.281\cdot10^{-7}$	 &$7.390\cdot10^{-9}$	 &$1.222\cdot10^{-7}$	\\
$14.2$	 &$1.919\cdot10^{-6}$	 &$1.994\cdot10^{-7}$	 &$3.053\cdot10^{-8}$	 &$9.983\cdot10^{-8}$	\\
$18.2$	 &$1.662\cdot10^{-6}$	 &$1.678\cdot10^{-7}$	 &$1.144\cdot10^{-8}$	 &$7.849\cdot10^{-8}$	\\
$23.3$	 &$1.449\cdot10^{-6}$	 &$1.455\cdot10^{-7}$	 &$3.317\cdot10^{-8}$	 &$6.319\cdot10^{-8}$	\\
$29.9$	 &$1.174\cdot10^{-6}$	 &$1.259\cdot10^{-7}$	 &$4.464\cdot10^{-8}$	 &$4.946\cdot10^{-8}$	\\
$38.2$	 &$9.886\cdot10^{-7}$	 &$1.070\cdot10^{-7}$	 &$2.264\cdot10^{-8}$	 &$4.004\cdot10^{-8}$	\\
$49.0$	 &$7.827\cdot10^{-7}$	 &$9.446\cdot10^{-8}$	 &$1.451\cdot10^{-9}$	 &$3.092\cdot10^{-8}$	\\
$62.7$	 &$6.077\cdot10^{-7}$	 &$8.263\cdot10^{-8}$	 &$3.772\cdot10^{-8}$	 &$2.559\cdot10^{-8}$	\\
$80.3$	 &$4.535\cdot10^{-7}$	 &$7.421\cdot10^{-8}$	 &$3.821\cdot10^{-8}$	 &$2.175\cdot10^{-8}$	\\
$102.8$	 &$3.844\cdot10^{-7}$	 &$6.813\cdot10^{-8}$	 &$4.051\cdot10^{-8}$	 &$1.840\cdot10^{-8}$	\\
$131.6$	 &$3.154\cdot10^{-7}$	 &$6.107\cdot10^{-8}$	 &$4.130\cdot10^{-8}$	 &$1.663\cdot10^{-8}$	\\
$168.6$	 &$2.728\cdot10^{-7}$	 &$5.770\cdot10^{-8}$	 &$2.486\cdot10^{-8}$	 &$1.701\cdot10^{-8}$	\\
$215.8$	 &$1.906\cdot10^{-7}$	 &$5.751\cdot10^{-8}$	 &$3.809\cdot10^{-8}$	 &$2.142\cdot10^{-8}$	\\
$276.4$	 &$1.338\cdot10^{-7}$	 &$8.237\cdot10^{-8}$	 &$3.853\cdot10^{-8}$	 &$6.264\cdot10^{-8}$	\\
\hline
\end{tabular}

\end{table}

\begin{table}

  \caption{The COSEBIs for $\vt_{\rm min} = 10$ arcsec and $\vt_{\rm
      max} = 100$ arcmin. The E-mode, $E_n$, and B-mode $B_n$,
    are given for the first five modes $n$, see
    Sect.~\ref{sec:results-EB}. The values
    of $\sigma$ indicate the error from the scaled Clone covariance diagonal
    (Sect.~\ref{sec:cov-sm}).}
  \label{tab:cosebis_100}

  \tabcolsep1ex
  \renewcommand{\arraystretch}{1}
\begin{tabular}{lllll}\hline\hline
\rule[-3mm]{0em}{8mm}$n$	 &$E_n$	 &$\sigma[E_n]$	 &$B_n$	 &$\sigma[B_n]$	\\\hline
\rule[-1mm]{0em}{5mm}$1$	 &$2.151\cdot10^{-10}$	 &$2.748\cdot10^{-11}$	 &$1.242\cdot10^{-11}$	 &$1.166\cdot10^{-11}$	\\
$2$	 &$2.288\cdot10^{-10}$	 &$4.814\cdot10^{-11}$	 &$1.706\cdot10^{-11}$	 &$2.195\cdot10^{-11}$	\\
$3$	 &$1.573\cdot10^{-10}$	 &$6.157\cdot10^{-11}$	 &$1.689\cdot10^{-11}$	 &$3.129\cdot10^{-11}$	\\
$4$	 &$1.368\cdot10^{-10}$	 &$6.765\cdot10^{-11}$	 &$-8.415\cdot10^{-12}$	 &$3.779\cdot10^{-11}$	\\
$5$	 &$1.557\cdot10^{-10}$	 &$6.736\cdot10^{-11}$	 &$-3.866\cdot10^{-11}$	 &$3.971\cdot10^{-11}$	\\
\hline
\end{tabular}

\end{table}

\begin{table}

  \caption{The COSEBIs for $\vt_{\rm min} = 10$ arcsec and $\vt_{\rm
      max} = 250$ arcmin. See Table \ref{tab:cosebis_100} for details.}
  \label{tab:cosebis_250}

  \tabcolsep1ex
  \renewcommand{\arraystretch}{1}
\begin{tabular}{lllll}\hline\hline
\rule[-3mm]{0em}{8mm}$n$	 &$E_n$	 &$\sigma[E_n]$	 &$B_n$	 &$\sigma[B_n]$	\\\hline
\rule[-1mm]{0em}{5mm}$1$	 &$4.841\cdot10^{-10}$	 &$1.469\cdot10^{-10}$	 &$9.532\cdot10^{-11}$	 &$1.156\cdot10^{-10}$	\\
$2$	 &$3.568\cdot10^{-10}$	 &$3.318\cdot10^{-10}$	 &$1.729\cdot10^{-10}$	 &$2.839\cdot10^{-10}$	\\
$3$	 &$-7.270\cdot10^{-11}$	 &$5.712\cdot10^{-10}$	 &$2.531\cdot10^{-10}$	 &$5.085\cdot10^{-10}$	\\
$4$	 &$-5.526\cdot10^{-10}$	 &$8.092\cdot10^{-10}$	 &$2.990\cdot10^{-10}$	 &$7.240\cdot10^{-10}$	\\
$5$	 &$-1.062\cdot10^{-9}$	 &$9.913\cdot10^{-10}$	 &$3.302\cdot10^{-10}$	 &$8.724\cdot10^{-10}$	\\
\hline
\end{tabular}

\end{table}

\end{appendix}

\end{document}